\documentclass[fleqn,usenatbib]{mnras}

\usepackage{newtxtext,newtxmath}

\usepackage[T1]{fontenc}

\DeclareRobustCommand{\VAN}[3]{#2}
\let\VANthebibliography\thebibliography
\def\thebibliography{\DeclareRobustCommand{\VAN}[3]{##3}\VANthebibliography}

\newcommand{\bm}[1]{\mbox{\boldmath$#1$}}

\usepackage{graphicx}	
\usepackage{amsmath}	
\usepackage{listings}
\usepackage{multicol}
\usepackage{xcolor}

\title[The  evolution of a non-compact  binary with misaligned spin and orbital angular momenta ]{On the non-dissipative orbital evolution of
a  binary system  comprising non-compact components with  misaligned spin and orbital angular momenta}

\author[Y. A. Lazovik, P. B. Ivanov \& J. C. B.
Papaloizou] 
{Y. A. Lazovik$^{1,2}$\thanks{E-mail: yaroslav.lazovik@gmail.com}
P. B. Ivanov$^{3}$\thanks{E-mail: pavel000astrophysics@gmail.com}
J. C. B. Papaloizou$^{4}$\thanks{E-mail: jcbp2@cam.ac.uk}
\\
$^{1}$Lomonosov Moscow State University, Faculty of Physics, 1 Leninskie Gory, bldg.2, Moscow, 119991, Russia\\
$^{2}$Sternberg Astronomical Institute, Lomonosov Moscow State University, Universitetsky pr. 13, Moscow, 119234, Russia\\
$^3$  P.N. Lebedev Physical Institute, 53 Leninsky Prospect, Moscow, 119991, Russia
\\
$^4$  DAMTP, Centre for Mathematical Sciences, University of Cambridge, Wilberforce Road, Cambridge CB3 0WA, UK}

\date{Accepted XXX. Received YYY; in original form ZZZ}

\pubyear{\the\year{}}

\begin{document}
\label{firstpage}
\pagerange{\pageref{firstpage}--\pageref{lastpage}}
\maketitle

\begin{abstract}
In this Paper we determine the non-dissipative tidal evolution of a close binary system with an arbitrary eccentricity in which the spin angular momenta of both components are misaligned with the orbital angular momentum. We focus on the situation where the orbital angular momentum dominates the spin angular momenta and so remains at small inclination to the conserved total angular momentum. Torques arising from rotational distortion and tidal distortion taking account of Coriolis forces are included. This extends the previous work of Ivanov \& Papaloizou relaxing the limitation resulting from the assumption that one of the components is compact and has zero spin angular momentum. Unlike the above study, the evolution of spin-orbit inclination angles is driven by both types of torque. We develop a simple analytic theory describing the evolution of orbital angles and compare it with direct numerical simulations. We find that the tidal torque prevails near  'critical curves' in parameter space where the time-averaged apsidal precession rate is close to zero. In the limit of small spin, these curves exist only for systems that have at least one component with retrograde rotation. As in our previous work,  we find solutions close to these curves for which the apsidal angle librates. As noted there, this could result in oscillation between prograde and retrograde states. We consider the application of our approach to systems with parameters similar to those of the misaligned binary DI Her. 
\end{abstract}




\section{Introduction}
Binary and exoplanetary systems may be such that the orientation
of the spins of one or both components does not coincide with that of the orbital angular momentum.
 Recently, this possibility has received observational
confirmation  \citep[see e.g.][]{Albrecht2, Marcussen}. 
In particular, \citet{Marcussen} find that increasing misalignment is correlated with increasing eccentricity (see also e.g. \citealt{Mart}, \citealt{Sh}, \citealt{Albrecht1}, \citealt{Philippov}, and \citealt{Liang} for a discussion of strong
misalignment in the well studied binary system DI Herculis). 
Exoplanetary systems can also contain  close-in planets on orbits with angular momentum vector strongly misaligned  with
respect to the rotational axis of the parent star 
\citep[see e.g.][]{Albrecht3, UM2022}.

For sufficiently small separation of the components of a binary/exoplanetary system, tidal interactions play a significant
role in determining orbital evolution \citep[see e.g.][and references therein]{Ogilvie2014}.
\citet{IP} (IP)  evaluated the tidal torques acting in a binary system with misaligned orbital and spin angular momenta
 assuming that one of the components is compact and therefore neglecting its spin angular momentum. 
The orbit was allowed to have arbitrary eccentricity. In particular, they took account of the effects of inertia and Coriolis forces on the tidal response which leads to a dependence of the torque on the orientation of the orbital line of apsides.
They found that tidal torques are present even when, as in this paper,  dissipative effects are not taken into account. Physically, this
can be explained by the action of the inertia and Coriolis forces, which results in  
the line joining the centre of masses of
the perturbed and perturbing body ceasing to be a symmetry axis associated with the tidal bulge. 
This gives rise to non-dissipative torques operating in the system, see also \cite{IP} and \cite{IP1}.

This was later extended by \citet{IP1} (IP1) to include all effects contributing to the rate of advance of the
apsidal line. In the above study, the analytic treatment of the equations for the evolution of the orbit was presented which indicated libration of the apsidal line close to critical curves on which the rate of advance of the apsidal line is zero and the spin-orbit inclination is stationary. Preliminary numerical simulations were carried out by \citet{IP2} (IP2).
  
It is the purpose of this Paper to extend the work of IP, IP1, and IP2 to allow for both components of the binary system to be non-compact and so contain spin angular momentum. In this context, we have in mind systems such as DI Her 
 which has non-compact stellar components and exhibits strong misalignments \citep[see e.g.][]{Sh, Philippov, Liang},
 and we relate our generic discussion to this case.
  
Because there are now participating spin vectors for both components, the dynamical system has more degrees of freedom,  and so the evolution is more complex in this case. However, by applying analytic and numerical approaches, we are able to discuss how the spin-orbit inclination angles change, extend the concept of a critical curve
and find solutions for which the apsidal angle librates when the system is close to such a critical curve.

The plan of this paper is as follows.
In Section \ref{derivation}, we introduce the quantities
characterising the binary system. These consist of the orbital angular momentum ${\bf L},$ the spin angular momenta
of the components ${\bf S}_1$ and ${\bf S}_2$,  the total angular momentum ${\bf J}= {\bf L}+{\bf S}_1+{\bf S}_2,$ and
angles defining the orientation of these vectors with respect to a Cartesian coordinate system with $z$ axis in the direction of 
${\bf J}$. Arbitrary orbital eccentricities are allowed. In Sections \ref{evspinvectors} - \ref{finaleq}, we derive equations governing the evolution of these quantities driven by the action of tidal torques.
This includes torques arising from inertial effects and Coriolis forces that were presented in IP and discussed in the context of a binary system with a compact component in IP1 and IP2.
 The evolution of these quantities depends on the orientation
of the apsidal line and an expression for its rate of change is given and discussed in Section \ref{apsidal}. In most situations of interest, $|{\bf S}_k| \ll |{\bf L}|$ ($ k=1,2$), which leads to the inclination, $i,$ between the orbital and total angular momentum
being very small. We adapt our governing equations to this limit in Section \ref{aqual}.
We then proceed to give an approximate analytic discussion of the governing equations in Sections \ref{approxdelta} - \ref{critical_curves},
providing estimates for the variation
of the spin-orbit inclination angles.  We also consider the evolution near critical curves, on which the rate of advance of the apsidal line is zero in a time-averaged sense,
finding a description in terms of a forced simple pendulum that is applicable in the limiting cases.
In Section \ref{critc}, we determine the location of the critical curves numerically for a range of semi-major axes, rotation rates of the binary components,
and spin-orbit inclination angles. The masses and radii of the components are specified as those appropriate to the DI Her system on which we particularly focus. The models are obtained from stellar evolution calculations using the \texttt{MESA} code as described in Appendix \ref{appendixA}.
Then, in Section \ref{orbitalcritc}, we determine the orbital evolution of representative systems numerically, successfully comparing the results with our analytic treatment.
 The numerical results show libration and circulation in the vicinity of critical curves as expected. In particular, this is the case for a model system with spin-orbit 
 inclination angles similar to that of the secondary in the DI Her system.
 Finally, we summarise our results and conclude in Section \ref{conclusions}.

\section{The dynamical equations governing the system}
\label{derivation}
\subsection{Definition of the  basic quantities characterizing the system, coordinate systems, and related quantities}
The spin angular momenta of the stars are denoted hereafter as $\textbf{S}_{1,2}$, while the orbital angular momentum is $\textbf{L}$. It is assumed that the total angular momentum
\begin{equation}
\textbf{J}= \textbf{L} +\sum_{k=1,2}\textbf{S}_k
\label{eq1}
\end{equation}
is conserved. $S_1$, $S_2$, $L$, and $J$ stand for the absolute values of the corresponding quantities,
while $\textbf{s}_1$, $\textbf{s}_2$, $\textbf{l}$, and $\textbf{j}$ are unit vectors in the respective 
directions.

In what follows we use two orthonormal bases of unit vectors. The first one ($\textbf{e}_{x}$, 
$\textbf{e}_{y}$, $\textbf{e}_{z}$) is chosen in such a way that the vector $\textbf{e}_{z}$ is parallel to $\textbf{J}$.  The spin and orbital angular momentum vectors can be expressed through the basis vectors as
\begin{equation}
\textbf{S}_{k}=S_{k}(\cos \delta_k \textbf{e}_z
+\sin \delta_{k}(\cos \nu_k \textbf{e}_x+\sin \nu_k
\textbf{e}_y))\hspace{3mm}{\rm for}\hspace{2mm} k=1,2    
\label{eq2}    
\end{equation}
and
\begin{equation}
\textbf{L}=L(\cos i \textbf{e}_z
+\sin i(\cos \alpha\textbf{e}_x+\sin \alpha
\textbf{e}_y)),
\label{eq3}    
\end{equation}
Here, the angles $\delta_{k}$ and $\nu_k$ are the angles between ${\textbf S}_k$ and ${\textbf J}$  and the azimuthal angle of ${\textbf S}_k$, respectively. The latter angle is measured in the plane containing $({\textbf e}_x, {\textbf e}_y)$  and relative to ${\textbf e}_x$. Similarly, $i$ and ${\alpha}$ are the angles between  ${\textbf L}$ and ${\textbf J}$ and the corresponding  azimuthal angle of  ${\textbf J}$, respectively.

The second set of orthonormal bases is associated with directions of the spin and orbital angular momenta. As discussed in e.g.  IP, IP1, and IP2, torques determining the non-dissipative orbital evolution are expected to be perpendicular to the directions of the spin vectors $\textbf{s}_k$. It is accordingly convenient to describe them with the help of the  orthonormal bases determined by the unit vectors
{\bf \begin{equation}
\textbf{s}_{\parallel, k}=\frac{(\textbf{l}-\cos \beta_k \textbf{s}_k)}{\sin \beta_k}, 
\quad 
 \textbf{s}_{\perp, k}=   \frac{\textbf{s}_k\times 
 \textbf{l}}{\sin \beta_k}, {\hspace{2mm}{\rm and}} \quad \textbf{s}_k, {\hspace{3mm} {\rm for}\hspace{2mm} k=1, 2,}
 \label{eq4}
\end{equation}}
where $\beta_k$ is the angle between the unit vectors 
$\textbf{s}_k$ and $\textbf{l}$. It can be found from  equations (\ref{eq2}) and (\ref{eq3}) to be given by
\begin{equation}
\cos \beta_k=(\textbf{s}_k\cdot \textbf{l})=
\cos\delta_k \cos i+\sin \delta_k \sin i \cos (\nu_k-\alpha) 
\label{eq5}
\end{equation}
Note that suffices  $\parallel$ and $\perp$ are assigned to vectors, which are parallel and perpendicular to the plane containing vectors $\textbf{s}_k$ and $\textbf{l}$, respectively.
 
From the identity
\begin{align}
{\textbf J}\cdot({\textbf L}+ \sum_k {\textbf S}_k)=J^2, \hspace{3mm}{\rm we\hspace{2mm} obtain}\label{j-1}
\end{align}
\begin{align}
L\cos i +\sum_k S_k\cos\delta_k=J\label{j0}
\end{align}
In addition, two useful relations follow from considering the $({\textbf e}_x, {\textbf e}_y, {\textbf e}_z)$  components of the identity
\begin{align}
{\textbf J}\times ({\textbf L}+ \sum_k {\textbf S}_k)=0.
\label{j1}
\end{align}
These yield the pair of equations
\begin{align}
&\sum_k S_k\sin\delta_k\sin(\alpha-\nu_k)=0 \hspace{3mm}{\rm and}
\label{j2}\\
&\sum_k S_k\sin\delta_k\cos(\alpha-\nu_k) +L\sin i=0.
\label{j3}
\end{align}
From (\ref{j2}) and (\ref{j3}), it follows that
\begin{align}
&\cos(\alpha-\nu_1)=\frac{S_2^2\sin^2\delta_2-S_1^2\sin^2\delta_1-L^2\sin^2i}{2S_1L\sin i\sin\delta_1},\label{j4}\\
&\cos(\alpha-\nu_2)=\frac{S_1^2\sin^2\delta_1-S_2^2\sin^2\delta_2-L^2\sin^2i}{2S_2L\sin i\sin\delta_2}\label{j5}\\
&S_1\sin(\alpha-\nu_1)= -S_2\frac{\sin\delta_2}{\sin\delta_1}\sin(\alpha-\nu_2)= \nonumber\\
&\pm\sqrt{\frac{4S_1^2S_2^2\sin^2\delta_1\sin^2\delta_2-(S_1^2\sin^2\delta_1+S_2^2\sin^2\delta_2-L^2\sin^2i)^2}
{4L^2\sin^2\delta_1\sin^2 i}}\label{j6}
\end{align}
Note that the positive sign of the square root is taken if $\alpha-\nu_1$ is in the interval $(0,\pi)$.
Otherwise, the negative sign is taken.
\subsection{Relation to the coordinate systems adopted in previous work {\bf and the angles determining the position 
of apsidal line}}
\label{subsec:rel}
In this work, we make use of the Cartesian coordinate system with axes in the direction of the basis vectors ${\bf e}_x, {\bf e}_y,$  and  ${\bf e}_z.$ The $Z$ axis is taken to  point
in the direction ${\bf e}_z$, being that of the total angular momentum ${\bf J}$. 

Similarly,  a coordinate system with the $Z'$ axis being co-directed with ${\bf L}$, as defined by (\ref{eq3}), can be chosen to correspond to the $(X',Y',Z')$ system  adopted by (IP, IP1, IP2). Clearly, the $(X',Y')$ plane is aligned with the orbital plane.

We also need to introduce coordinate systems with $Z^{''}_k$ axes pointing in the direction ${\bf S}_k,  k =1,2$,
with the  $ (X^{''}_k,Y^{''}_k)$ planes aligned with the equatorial planes of the respective components.
In (IP, IP1, IP2), this was set up for only one component, which led to the definition of the $(X,Y,Z)$ system. Note a difference in notations between this work and IP. While in this Paper we use double primed values $(X^{''}_k,Y^{''}_k,Z^{''}_k)$ to characterise the coordinates associated with stellar spins, in IP, this notation was used to characterise the coordinate system associated with the total angular momentum, and vice versa. This
difference is due to the fact that, in this Paper, we formulate our basic equations in the coordinate system associated with the total angular momentum. On the contrary, in IP, the main role was played by the coordinate system associated with stellar spin.   

For each of these systems the axis $OY^{''}_k$ is directed along the vector $\textbf{s}_{\perp, k}$ defined in eq. (\ref{eq4}),
while
the direction of the axis  $OY$ in the $(X, Y)$ plane can be defined through an analogous vector, but 
determined with the help of the unit vector ${\bf j}$
\begin{align}
\textbf{j}_{\perp} = \frac{ {\bf j}\times {\bf l}}{\sin i}.\label{oy''}
\end{align}
Note that the vector $\textbf{s}_{\perp, k}$ lies in the orbital plane as does the corresponding vector $OY$  adopted in (IP,IP1,IP2).
The vector $\textbf{j}_{\perp}$   defined here, being perpendicular to both ${\bf l}$ and ${\bf j},$  also lies in the orbital plane
and can be used to define the line of nodes.

The argument of pericentre or apsidal angle,  $\varpi$ is measured in the orbital plane and we can choose to do this with respect to the line $OY$.
Though note that  IP effectively allowed for an arbitrary rotation of the reference line with respect to the line of nodes, through the angle $\gamma$,  which has no impact in the case  considered there, where only one component
is involved and which may be taken to be zero (see also below). 
 
When calculating tidal torque on the $k {\rm th}$  component, following  IP, the argument of pericenter has to be measured with respect
 to $OY^{''}_k.$ This is defined to be, $ \hat {\varpi_k} = \varpi +\gamma_k,$ where $\gamma_k$ is the angle between 
$OY$ and $OY^{''}_k$ (or, in other words, between  $\textbf{j}_{\perp}$ and $\textbf{s}_{\perp, k}$). 
We take this to increase as we move from $OY^{''}_k$ to $OY$ corresponding to a right-handed rotation. Making use of (\ref{eq4}) and (\ref{oy''}), we may write
\begin{align}
&\cos\gamma_k = \frac{ ({\bf s}_k\times {\bf l})\cdot({\bf j}\times {\bf l})}{\sin\beta_k\sin i},\quad {\rm and}\nonumber\\
&\sin\gamma_k =  \frac{( ({\bf s}_k\times {\bf l})\times({\bf j}\times {\bf l}))\cdot {\bf l}}{\sin\beta_k\sin i}.\label{gammak}
\end{align}
From (\ref{gammak}),  (\ref{eq2}), and (\ref{eq3})  we readily obtain
\begin{align}
&\cos\gamma_k= \frac{\cos\delta_k -\cos\beta_k\cos i}{\sin\delta_k\sin i}=-\cos i\cos(\alpha-\nu_k)+\frac{\sin i\cos\delta_k}{\sin\delta_k}, \nonumber\\
&\sin\gamma_k=-\frac{\sin\delta_k\sin(\alpha-\nu_k)}{\sin\beta_k}.\label{gammak1}
\end{align}
 
In the limiting case, for which the secondary becomes compact and ${ S}_2=0$, we find from eqns (\ref{j2})-(\ref{j6}) that $\alpha -\nu_1 =\pi$ 
 and $L\sin i=S_1\sin\delta_1,$
with $\alpha-\nu_2$ being indeterminate.
 In addition, eq.  (\ref{j-1}) yields $L\cos i+S_1\cos\delta_1=J.$
 As the unit vectors ${\bf j}, {\bf l}, $ and ${\bf s}_1$ are coplanar in this limit,  $\textbf{s}_{\perp, 1}$ 
and $\textbf{j}_{\perp}$ are co-linear corresponding to $\gamma_1=0.$ 
 
For some purposes, it may be useful to introduce the longitude of periapsis $\Pi$. This is defined as, $\varpi+ \Omega$, where $\Omega$ here
is the longitude of the node. This is the angle between $OY$ and a line in the $(X,Y)$ plane that remains fixed in an inertial frame.
When $i=0,$ this angle is effectively $\varpi$ measured with respect to a line that is fixed in an inertial frame, rather than with respect to $OY$
that may be moving rapidly on account of the precession of the line of nodes.

\subsection{Graphical representations of the angles and coordinate systems adopted in the present study}

\begin{figure*}
\begin{multicols}{2}
    \includegraphics[width=\linewidth]{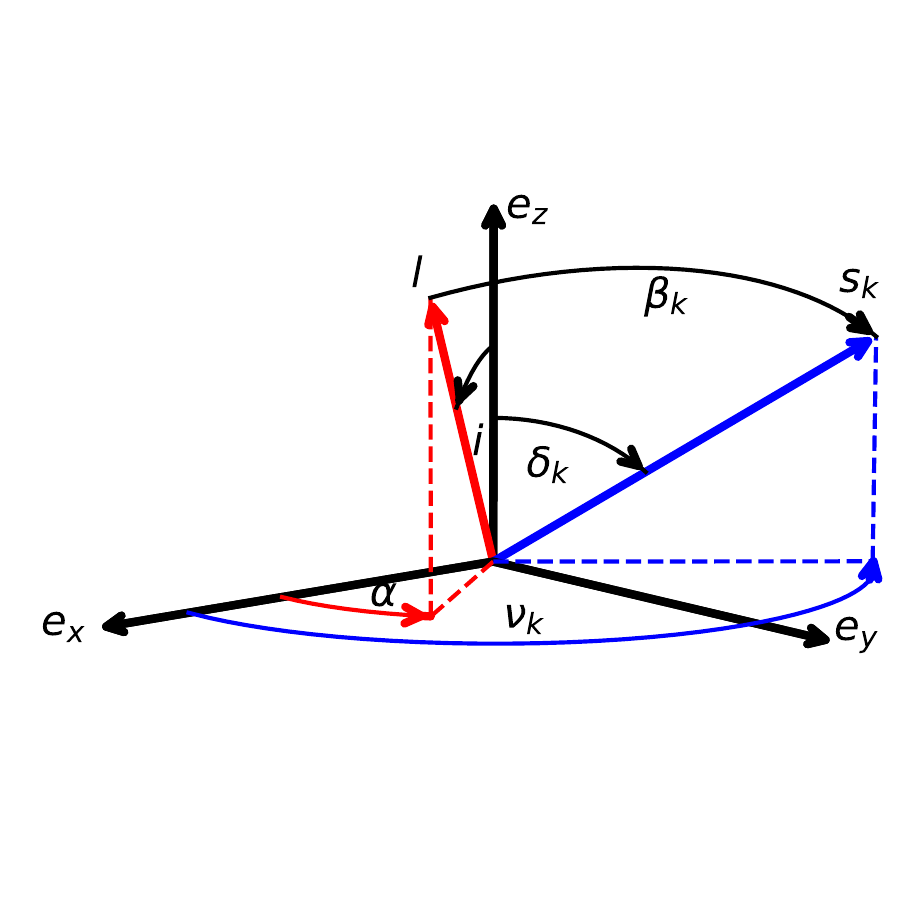}\par 
    \includegraphics[width=\linewidth]{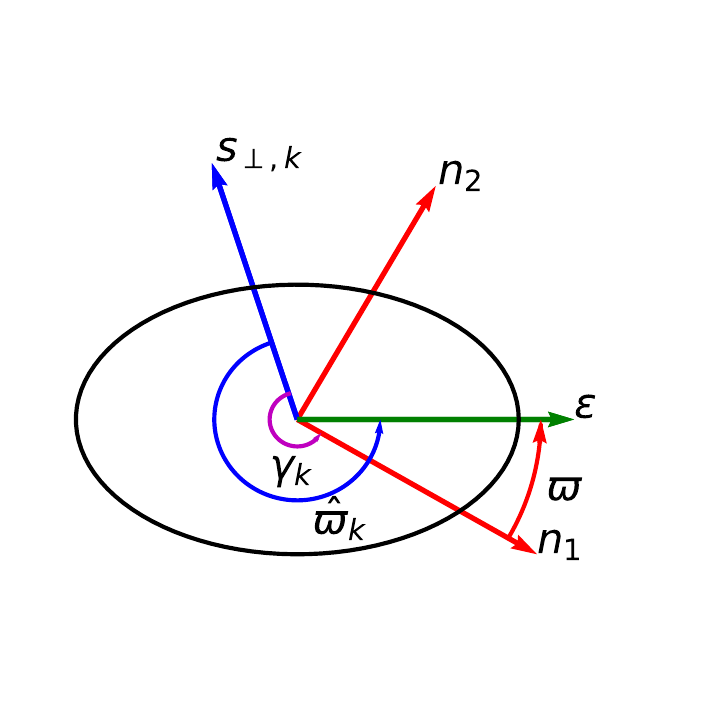}\par 
    \end{multicols}
    \includegraphics[scale=0.6]{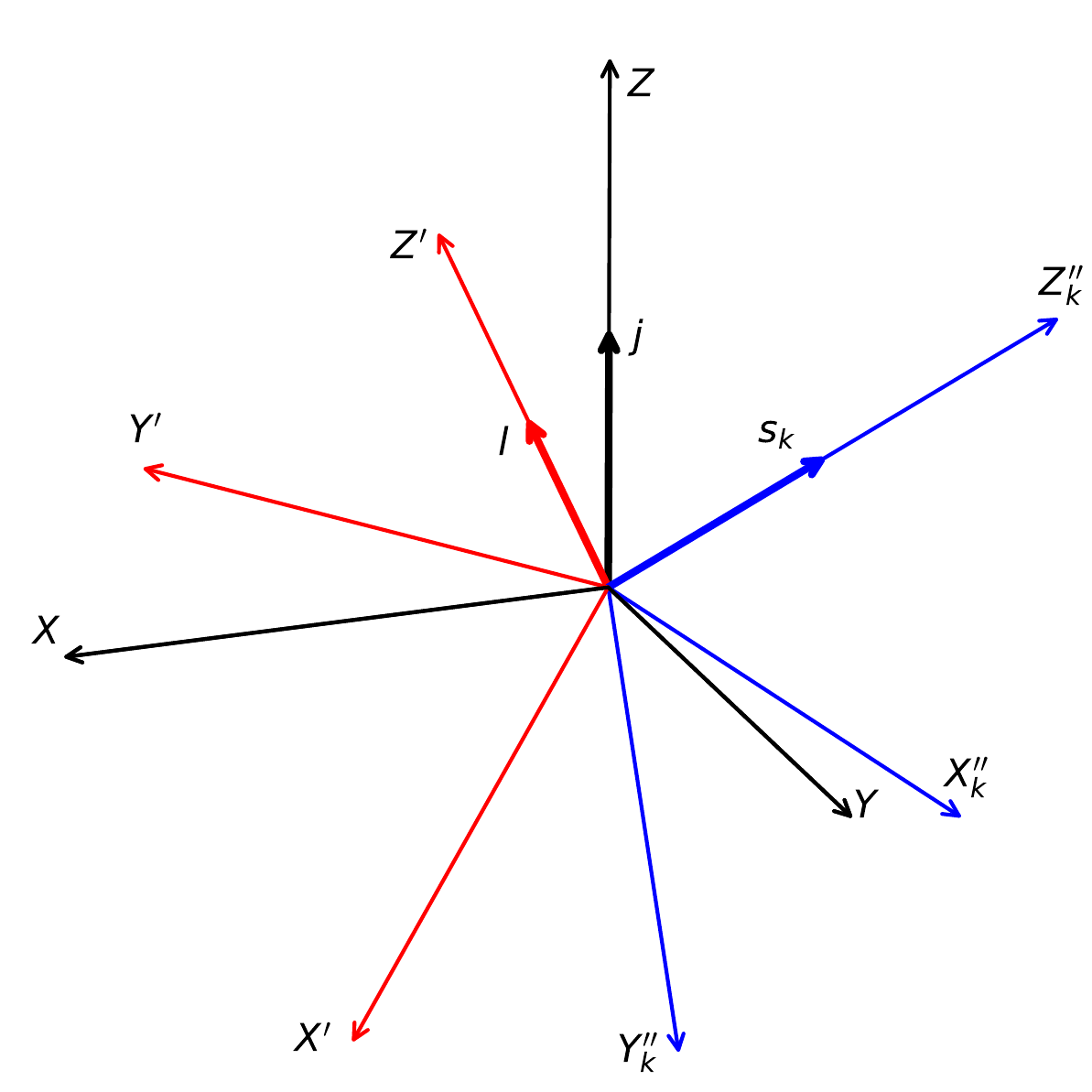}\par
\caption{Top left panel. We show the inclination angles $\delta_k$ between the unit vector ${\bf e}_z$ pointing in the direction 
of the total angular momentum and the unit vector in the direction of the spin vector of a particular component,
${\bf s}_k$, the inclination angle $i$ between  ${\bf e}_z$ and the unit vector in the direction of the orbital angular
momentum, ${\bf l}$, as well as the angle $\beta_k$ between  ${\bf l}$ and  ${\bf s}_k$. The 'rotational' angles $\nu_k$ and
$\alpha$ define the position of projections of  ${\bf s}_k$ and  ${\bf l}$ onto the plane perpendicular to  ${\bf e}_z$.
Top right panel. The angles $\hat \varpi_k$, $\varpi$, and $\gamma_k$ are shown in the orbital plane. The ellipse shows the orbit and the unit vector ${\bf \epsilon}$ is directed towards the periastron. The unit vectors ${\bf s}_{\perp,k}$ and ${\bf n}_1\equiv \textbf{j}_{\perp}$ point in the directions of the nodal lines corresponding to the planes perpendicular to the spin
angular momentum of a particular component and the total angular momentum, respectively. These vectors and ${\bf n}_2$ lie in the orbital plane. {Bottom panel. We illustrate the $(X,Y,Z)$, $(X',Y',Z')$, and $(X^{''}_k,Y^{''}_k,Z^{''}_k)$ coordinate systems, adopted in the present study. Note that the $Y, Y'$, and $Y^{''}_k$ axes are coplanar.} }   
\label{Plots}
\end{figure*}

In a schematic plot shown on the top left 
panel of Fig. \ref{Plots}, we show the angles $\alpha$, $\beta_{k}$, $\delta_{k}$, $i$, and $\nu_k$, defined in the relations (\ref{eq2}), (\ref{e3}), and
(\ref{eq5}). The top right panel schematically shows the angles $\hat \varpi_k$, $\varpi$, and $\gamma_k$, defined through
eq. (\ref{gammak}), in the orbital plane. Note that ${\bf n}_1\equiv \textbf{j}_{\perp}$, while ${\bf n}_1$ is perpendicular
to  ${\bf n}_2$, and the vectors  ${\bf n}_1, {\bf n}_2$ form a right-handed orthogonal pair in the orbital plane, see eqns (\ref{en5}) below. {The bottom panel of Fig. \ref{Plots} demonstrates the $(X,Y,Z)$, $(X',Y',Z')$, and $(X^{''}_k,Y^{''}_k,Z^{''}_k)$ coordinate systems together with the unit vectors ${\bf j}$, ${\bf l}$, and ${\bf s_k}$, defining the orientations of the $Z, Z'$, and $Z^{''}_k$ axes, respectively. Note that, unlike in IP, IP1, and IP2, these axes are not coplanar. At the same time, the $Y, Y'$, and $Y^{''}_k$ axes lie in the orbital plane.} 

\subsection{Evolution of the spin vectors}\label{evspinvectors}
The spin vectors evolve due to the action of torques induced by the tidal interaction. As indicated above, when considering conservative evolution, the torque acting on ${\textbf S}_k$ is assumed to be perpendicular to ${\textbf S}_k$ and so is given by the generic equation 
\begin{equation}
\frac{d\textbf{S}_{k}}{dt}=\textbf{T}_{k}, {\hspace{2mm}{\rm with}} \quad \textbf{T}_{k}
=T_{\parallel, k}\textbf{s}_{\parallel, k}+
T_{\perp, k}\textbf{s}_{\perp, k},
\label{eq6}    
\end{equation}
where the scalar quantities $T_{\parallel, k}$ and $T_{\perp, k}$ can be found from an analysis of the tidal interaction. The rate of evolution of orbital angular momentum follows 
from total angular momentum conservation in the form
\begin{equation}
\frac{d\textbf{L}}{dt}=-\sum_{k=1,2}\textbf{T}_k.
 \label{eq7}   
\end{equation}
It is useful to have explicit expressions for the components of the torque in the basis $({\bf e}_x, {\bf e}_y, {\bf e}_z),$ associated with $\textbf{J}$. These can be found by  making use of eqns (\ref{eq2}) and (\ref{eq3})
to substitute for ${\bf l}$ and  ${\bf s}_k$ in eq. (\ref{eq4}) in the form $T_{\parallel, k}\textbf{s}_{\parallel, k}=  (T^{x}_{\parallel, k},T^{y}_{\parallel, k},T^{z}_{\parallel, k})$, where
\begin{align}
&T^{z}_{\parallel, k}=\frac{T_{\parallel, k}}{\sin \beta_k}(\cos i
-\cos \beta_k\cos \delta_k),
\label{eq8}\\      
&T^{x}_{\parallel, k}=\frac{T_{\parallel, k}}{\sin \beta_k}(\sin i \cos \alpha
-\cos \beta_k\sin \delta_k\cos \nu_k), \label{eq9}\\
\quad
&T^{y}_{\parallel, k}=\frac{T_{\parallel, k}}{\sin \beta_k}(\sin i \sin \alpha
-\cos \beta_k\sin \delta_k\sin \nu_k), 
\label{eq10}
\end{align}
together with $T_{\perp, k}\textbf{s}_{\perp, k}=  (T^{x}_{\perp, k},T^{y}_{\perp, k},T^{z}_{\perp, k})$, where
\begin{align}
&
T^{z}_{\perp, k}=\frac{T_{\perp, k}}{\sin \beta_k}
\sin \delta_k \sin i \sin (\alpha-\nu_k),
\label{eq11}  \\
&T^{x}_{\perp, k}=\frac{T_{\perp, k}}{\sin \beta_k}
(\sin \delta_k \cos i \sin \nu_k -\cos \delta_k 
\sin i \sin \alpha), \quad \label{eq12}
\\
&T^{y}_{\perp, k}=\frac{T_{\perp, k}}{\sin \beta_k}
(\cos \delta_k \sin  i \cos \alpha -\sin \delta_k 
\cos i \cos \nu_k).
\label{eq13}    
\end{align}
\subsection{Equations governing the  evolution  of the inclination and azimuthal angles}\label{evangles}
The evolution equation for the absolute value of orbital angular momentum, $L$, can be easily obtained by differentiating the identity $L^2=\textbf{L}\cdot \textbf{L}$ over time and using (\ref{eq7}). We have
\begin{equation}
\frac{dL}{dt}=-\sum_k T^{k}_{\parallel}\sin \beta_k.
\label{eq14}    
\end{equation}
A differential equation for the evolution of the inclination angle $i$ is obtained from the $z$ component of (\ref{eq7}) in the form
\begin{equation}
\begin{split}
\frac{di}{dt}= \frac{1}{L\sin i}
\sum_k \frac{1}{\sin \beta_k} \biggl(
  &T^{k}_{\parallel}\cos\beta_k(\cos\beta_k\cos i -\cos \delta_k) + {}\\
  & T^{k}_{\perp}\sin i\sin \delta_k \sin (\alpha -\nu_k)
  \biggr)
\end{split}
\label{eq15} 
\end{equation}

The evolution equation for the azimuthal angle $\alpha$ can be obtained from a linear combination of the $x$ and $y$ components of eq. (\ref{eq7}) with the result

\begin{equation}
\begin{split}
\frac{d\alpha}{dt}=-\frac{1}{L\sin i} \biggl(
  &\sum_k \frac{1}{\sin \beta_k}(T^{k}_{\parallel}\cos \beta_k \sin \delta_k \sin (\alpha - \nu_k) + {}\\
  & T^{k}_{\perp}(\sin i
\cos \delta_k -\cos i \sin \delta_k\cos (\alpha -\nu_k))
  \biggr)
\end{split}
\label{eq16} 
\end{equation}

The evolution equations for the inclination and azimuthal angles associated with the stellar spins,
$\delta_k$ and $\nu_k$ are derived in a similar way, though now making use of (\ref{eq6}) together with taking into account that the magnitudes of the stellar spins, $S_k,$ ($k=1,2$)  remain constant. We thus obtain
\begin{equation}
\begin{split}
\frac{d\delta_k}{dt}= &-\frac{T^{k}_{\parallel}}{S_{k}\sin \beta_k \sin \delta_k} (\cos i -\cos \beta_k \cos \delta_{k}) - {} \\
 & \frac{T^{k}_{\perp}}{S_k\sin \beta_k}\sin i \sin (\alpha - \nu_k),\quad{\rm and}
\end{split}
\label{eq17} 
\end{equation}

\begin{equation}
\begin{split}
\frac{d\nu_k}{dt}=&\frac{T^{k}_{\parallel}}{S_k {\sin \delta_k} \sin \beta_{k}}\sin i \sin (\alpha - \nu_k)+ {} \\
 & \frac{T^{k}_{\perp}}{S_k \sin  \delta_k \sin \beta_{k}}(\sin i \cos \delta_{k}\cos(\alpha - \nu_k) -\cos i \sin \delta_{k}).
\end{split}
\label{eq18} 
\end{equation}

\subsection{Reduction to the final set of governing equations}\label{finaleq}
{Eqns (\ref{eq14}) - (\ref{eq18})  provide seven equations for the evolution of $L, i,\alpha,$ $\nu_k,$ and $\delta_k,$ for $k=1,2.$ We note that $\alpha$ and $\nu_k$ occur only in the form  $\alpha-\nu_k $ on the right-hand sides. Thus, these quantities can be solved for, leaving   $\alpha$  to be separately determined from eq. (\ref{eq16}) after they and 
the other quantities have been determined. 
 Furthermore, eq. (\ref{j0}) and eqns (\ref{j4}) - (\ref{j6}) can be used to specify $L$ and  $\alpha -\nu_k$, enabling (\ref{eq14}) and (\ref{eq18}) to be discarded. {Eqns (\ref{eq15}) and (\ref{eq17}) provide  a total of three equations for $i$ and $\delta_k,$ for $k=1,2.$ 
These are supplemented with an equation for $d\varpi/dt$, which is required to specify the torques, leading to four equations in total
governing the evolution of these quantities.
We specify the equation for $d\varpi/dt$  below.}

\subsection{The evolution of the apsidal angle }\label{apsidal}
The equation governing the evolution of the angle 
$\varpi$ has the general form (see \cite{IP1}
and \cite{IP2})
\begin{align}
\frac{d{\varpi}}{dt}=\dot \varpi_\mathrm{T} +\dot \varpi_\mathrm{E} + \dot \varpi_\mathrm{R} + \dot \varpi_\mathrm{NI}.
\label{prec}
\end{align}     
In our previous work, the contributions to the apsidal angle evolution rate were
evaluated under the assumption that one of the components is compact
and therefore has zero spin angular momentum or tidal torques acting on it.
Here we generalise (\ref{prec}) to allow both components to be treated on an equal basis.
\subsubsection{The effect of tidal distortion}
Here, each component produces a tidal distortion of its companion, to the order we working,
acting as a point mass as it does so.   
On account of this, the contributions of each to  ${d \varpi_\mathrm{T} / dt}$ will be additive. Thus 
the distortion of the component labelled $k$ for $k=1,2$  contributes \citep[see e.g.][]{St1939}
\begin{align}
\left({\dot \varpi}_\mathrm{T} \right)
_k =15 k_{2,k}n_0\frac{M_{j}R_{*,k}^5}{M_{k}(a(1-e^2))^5} \left(1+\frac{3}{2} e^2
+\frac{1}{8}e^4\right),\label{Apse}
\end{align} 
 where $n_0$ is the mean motion.  
 Here we adopt the notation of IP1 and IP2 and imply that $j=1,2\ne k$ to denote quantities related to
the perturbing component.
 In total, we have
 \begin{align}
{\dot \varpi_\mathrm{T}} =\frac{15n_0} {(a(1-e^2))^5}\left(1+\frac{3}{2} e^2
+\frac{1}{8}e^4\right)
 {\displaystyle{\sum_{k=1,2}}} \frac{k_{2,k}{M_{j}R_{*,k}^5}}{M_k} ,\label{Apse1}
\end{align} 
 \subsubsection{Einstein precession}
 This is independent of the mass distribution of the components to the order we are working
 and so is of the same form as in IP1 and IP2. Thus we adopt
\begin{align}
{\dot \varpi_\mathrm{E}}
 =\frac{3G(M_1+M_2)}{c^2a(1-e^2)}n_{0}, \label{Ein}
\end{align}
which is the standard expression for the Einstein relativistic apsidal precession and $c$ is the speed of light.

\subsubsection{Orbital precession resulting from rotational distortion and non-inertial effects due to  spin precession}
These effects were considered for a system with a compact component in IP1 and IP2.
This is generalised here to treat the two components on an equal footing in   
Appendix \ref{app2}. The orbital precession rate resulting from rotational distortion together with that due to non-inertial effects
produced by the precession of the stellar spins is given by
eq. (\ref{a15}),
which can be written in the form

\begin{equation}
\begin{split}
\dot \varpi_\mathrm{R} +\dot\varpi_\mathrm{NI} =&- {\displaystyle{\sum_{k=1,2}}}\Omega_{Q,k} \biggl(  \frac{3\cos^2\beta_k -1 }{2} + {}\\
  &\cot i \cos\beta_k\frac{(\cos\delta_k-  \cos i  \cos\beta_k)}{\sin i}
  \biggr)
\end{split}
\label{NI0} 
\end{equation}

{A potentially useful equivalent form  for $\dot \varpi_\mathrm{R} +\dot\varpi_\mathrm{NI}$ is obtained if  eq. (\ref{a16}) is adopted as an alternative to eq. (\ref{a15}).   This yields}

\begin{equation}
\begin{split}
&\dot \varpi_\mathrm{R} +\dot\varpi_\mathrm{NI} =- {\displaystyle{\sum_{k=1,2}}}\Omega_{Q,k} \Biggl(  \frac{3\cos^2\beta_k -1 }{2} + {}\\
  &\frac{\cos i}{\sin^2 i}\cos\beta_k\biggl(
\left( \frac{L}{J}-\cos i\right)
\cos\beta_k +
{\displaystyle \sum_{j=1,2}}
\frac{S_j{\bf s}_j\cdot{\bf s}_k}{J} 
\biggr)
  \Biggr)
\end{split}
\label{NI} 
\end{equation}

We remark that the scalar product ${\bf s}_i\cdot{\bf s}_j$  can be found with the help of eq. (\ref{eq2}).
Thus, eq. (\ref{prec}) determines the evolution of ${\varpi}$ with eqns (\ref{Apse1})-(\ref{NI}) specifying ${\dot \varpi}_\mathrm{T},  {\dot \varpi}_\mathrm{E},$ and ${\dot \varpi}_\mathrm{R} +{\dot \varpi}_\mathrm{NI}$
therein.		
\subsection{The complete set of governing equations for \texorpdfstring{$i, \delta_k, k=1,2$ and $\varpi.$}{Lg}}
{These are eq. (\ref{eq15}) and the pair of eqns (\ref{eq17}), making use of the relations specified by 
eqns (\ref{j4})- (\ref{j6}), which can be used to remove the dependence on $\alpha -\nu_k$, together 
with eq. (\ref{prec}). They require the specification}
of  the torque components  $T_{\parallel}^k,$  and $T_{\perp}^k$.
These were determined in IP and are provided here in a suitable form in Appendix \ref{app1}.

\subsection{The case of a compact component}
{We now make the simplifying assumption that the size (and hence the spin) of the secondary component is negligible, implying $S_2 = 0$ and $S = S_1$. We have already referred to this case in Section~\ref{subsec:rel}. Below, we will take the limit of the equations governing the evolution of the orbital parameters, namely eqns (\ref{eq15} ) (\ref{eq17}), and (\ref{prec}), 
 as the secondary component becomes compact.}

{Noting, as remarked at the end of Section \ref{subsec:rel}  that, in this limit $\alpha - \nu_1 = \pi$, eq. (\ref{eq17}) becomes
\begin{equation}
\frac{d\delta}{dt}=-\frac{T_{\parallel}}{S \sin \beta \sin \delta} (\cos i -\cos \beta \cos \delta).
\label{eqL_1}  \\   
\end{equation}
Here, as only the primary component for which $k=1$ needs to be taken into account,  the subscript $k$, which becomes redundant,  is omitted. All the appearing angles 
apply to the primary component, the only one that contains angular momentum and can have non-zero torques acting on it. }

Substituting $i = \beta - \delta$ (which is valid because ${\bf{S}}$, ${\bf{L}}$, and ${\bf{J}}$ lie in the same plane in this limit) the above equation simplifies to become
\begin{equation}
\frac{d\delta}{dt}= - \frac{T_{\parallel}}{S},
\label{eqL_2}  \\   
\end{equation}
which is in agreement with eq. (10) of IP1 as the direction of the X-axis in the latter study and the direction of $\bf{t_{\parallel}}$ in the present study are the same. 

Similarly, in the limit in which the secondary is compact, eq. (\ref{eq15}) becomes
\begin{equation}
\frac{di}{dt}= - \frac{T_{\parallel}\cos \beta}{L}.
\label{eqL_3}
\end{equation}
Accordingly,
\begin{equation}
\frac{d\beta}{dt}=\frac{d\delta}{dt} + \frac{di}{dt} = -\left(\frac{1}{S} + \frac{\cos \beta}{L} \right) T_{\parallel},
\label{eqL_4}
\end{equation}
which is equivalent to eq. (12) of IP1. Finally, eqns (\ref{eq16}) and (\ref{eq18})  reduce to the expressions:
\begin{align}
&\frac{d\alpha}{dt} =  -\frac{T_{\perp}}{L \sin i}\label{eqL_5},\\
&\frac{d\nu}{dt}= -\frac{T_{\perp}}{S \sin \delta}.
\label{eqL_6}
\end{align}
{Thus, eqns (\ref{eqL_2})
 - (\ref{eqL_6}) represent a simplified form of eqns (\ref{eq15})--(\ref{eq18}) for the case of a binary system with a compact secondary component.}
\section{A qualitative analysis of the dynamical system in the limit when spin angular momenta are small
in comparison with the orbital angular momentum} \label{aqual}

In general, when the mean motion, $n_0$, is much smaller than the characteristic frequency $\Omega_{*}=\sqrt{GM/ R^3}$, 
 tidal interaction can be considered to be weak. When the condition\\ $n_0\ll \Omega_*$ holds,
the term determined by stellar flattening in the expression for the 'perpendicular' torque $T_{\perp,k}$, that is the second term on the right-hand side of (\ref{Tperp}), here denoted by $T_{\perp,k}^{s.f.},$
 is expected to be much larger in magnitude than the first term as well as $T_{\parallel,k}$.  This is the case  unless the stars are in  nearly polar orbits with
$\beta_k \approx \pi/2$. 

On the other hand, when $S_k \ll L$, $L\approx J$ and the inclination angle $i$ is small. In this situation, the
following approximation scheme is possible. \\
1) We neglect $T_{\parallel,k}$ and the first term on the right-hand side of (\ref{Tperp}) in comparison
to the second term,  
 $T^{s.f.}_{\perp,k}$, when these terms can be directly compared.\\
2) We assume, however,
that  $iT^{s.f.}_{\perp,k}$ can be of the order of $T_{\parallel,k}$. \\
3) When $i$ is small, from eq. (\ref{eq5}) we find  that 
$\delta_k\approx \beta_k$, and we can write  $\beta_k = \delta_k+\kappa_k,$ with correction to the first order in $i$
\begin{equation}
\kappa_k=-i\cos (\nu_k-\alpha).
\label{neq0}
\end{equation}

Adopting these approximations, we have from eq. (\ref{eq17})
\begin{equation}
{d\delta_k \over dt}=-{T_{\parallel,k}\over S_k}+{iT_{\perp,k}\over S_k\sin \delta_k}\sin(\nu_k-\alpha), 
\label{neq1}
\end{equation}
while eq. (\ref{eq18}) can be brought in
a very simple form
\begin{equation}
{d\nu_k \over dt}=-{T_{\perp,k}\over S_k\sin \delta_k},
\label{neq2}    
\end{equation}
and it is implied that $\beta_k=\delta_k$ in the expressions for $T_{\perp,k}$ and $T_{\parallel, k}$.

In order to find $\alpha$, we use (\ref{j2}), which allows us to express it in terms of $\nu_k$ as
\begin{equation}
\sin \alpha =-{(\sin \nu_1+\xi \sin \nu_2)\over \sqrt{1+\xi^2+2\xi \cos (\nu_1-\nu_2)}},
\quad \cos \alpha =-{(\cos \nu_1+\xi \cos \nu_2)\over \sqrt{1+\xi^2+2\xi \cos (\nu_1-\nu_2)}},
\label{neq3}    
\end{equation}
where $\xi={S_2\sin \delta_2 /( S_1 \sin \delta_1)}$, while (\ref{j3}) can be used to express
$i$ in terms of $\nu_k$ as
\begin{equation}
i={S_1\sin \delta_1 \over J}\sqrt{1+\xi^2+2\xi \cos (\nu_1-\nu_2)},
\label{neq4}
\end{equation}
and we note that  taking the square root  to be positive,  we chose the signs of the right-hand sides in (\ref{neq3}) to ensure
that $i$ is positive.  

We substitute (\ref{neq3}) and (\ref{neq4}) in (\ref{neq1}) to obtain
\begin{equation}
{d\delta_k \over dt}=-{T_{\parallel,k}\over S_k}-{S_j\sin \delta_j \over J S_k\sin \delta_k}T_{\perp,k}\sin(\nu_k-\nu_j),
\label{neq5}
\end{equation}
where $j \ne k$.

Note that we can set $L=J$ in eqns (\ref{neq1})--(\ref{neq5}) and assume that the orbital eccentricity $e$ is constant.

 We recall that $T_{\parallel,k}$ depends on $\hat \varpi=\varpi+\gamma_k$,  with $\gamma_k$ specified through equation (\ref{gammak1}). In  the limit $ i \rightarrow 0$, we can evaluate $\gamma_k$ as
\begin{align}
\gamma_k=\alpha-\nu_k+\pi,
\label{nprec0}
\end{align}    
while eq. (\ref{prec}) for the evolution of apsidal angle $\varpi $ takes the form
\begin{align}
\frac{d{\varpi}}{dt}=\dot \varpi_{*} + \dot \varpi_\mathrm{NI},
\label{nprec}
\end{align}  
 where $\dot \varpi_*=\dot \varpi_\mathrm{T} +\dot \varpi_\mathrm{E}+\dot \varpi_\mathrm{R}$ can be considered constant.   Note that $\dot \varpi_\mathrm{NI}$ is given  in Appendix \ref{app2},  
and  $\dot \varpi_\mathrm{R}$ does not contribute  to it  (see also
IP1).
{ From Appendix \ref{app2} we have}
\begin{align}
\dot \varpi_\mathrm{NI}\approx {1\over i}\sum_{k=1,2}\Omega_{Q,k}\sin \delta_k \cos \delta_k \cos (\nu_k-\alpha).
\label{nprec1}
\end{align}  
Now we substitute (\ref{neq3}) and (\ref{neq4}) to (\ref{nprec1}) to have
\begin{equation}
\begin{split}
\dot \varpi_\mathrm{NI}\approx&-J\Biggl(\left({B_1\over S_1}{(1+\xi \cos (\nu_1-\nu_2))\over (1+\xi^2+2\xi \cos (\nu_1-\nu_2)}\right) + {}\\
  &\left({B_2\over S_2}{(1+\xi^{'} \cos (\nu_1-\nu_2))\over (1+{\xi^{'}}^2+2\xi^{'} \cos (\nu_1-\nu_2))}\right)
  \Biggr),
\end{split}
\label{nprec2} 
\end{equation}

where $B_{k}=\cos \delta_{k}\Omega_{Q,k}$ and $\xi^{'}=\xi^{-1}={S_1\sin \delta_1/ S_2\sin \delta_2}$.

Eqns (\ref{neq2})--(\ref{nprec1} or \ref{nprec2}) form a complete set.

\subsection{Approximate solution in case when variations of \texorpdfstring{$\delta_k$}{Lg} are small}\label{approxdelta}

We  set $\delta_k=\delta_k^{0}+\Delta_k$ with $|\Delta_k| \ll |\delta^{0}_{k}|$ and make the approximation that $\delta_k=\delta_k^0$ in 
the right-hand sides of (\ref{neq2})--(\ref{nprec1}). We also assume that, without loss of generality,  when $t=0,$ $\Delta_k=0.$ 

Eq. (\ref{neq2}) implies that, in this case, $\nu_k$ are just linear functions of time.
Thus,
\begin{equation}
\nu_k=\omega_kt+\nu_k^0, \quad \omega_k=-{T_{\perp,k}\over S_k\sin \delta^0_k}.
 \label{neq7}   
\end{equation} 
Here we note that $\beta_k$ has been replaced by  $\delta_k$ which in turn is replaced by $\delta_k^0$ in $T_{\perp,k}^{s.f.},$ with
 $T_{\perp.k} $ approximated as  the latter.\\
The time dependencies of $\alpha$ and $i$ then immediately follow from (\ref{neq3}) and (\ref{neq4}).

We now set $\delta_k=\delta_k^0$ in r.h.s. of (\ref{nprec2}) and integrate (\ref{nprec}) with the result
\begin{equation}
\begin{split}
&\varpi=\varpi_0+\omega_{*}t - J\Biggl(\left({B_1\over S_1}+{B_2\over S_2}\right)\frac{ t}{2}+ {}\\
  &{1\over \Delta \omega}
\left({B_1\over S_1}-{B_2\over S_2}\right)\left ( \arctan {|1-\xi|\over 1+\xi}\tan
{x\over 2}-\arctan {|1-\xi|\over 1+\xi}\tan
{x^0\over 2}\right)\Biggr),
\end{split}
\label{neq8} 
\end{equation}

where we use (\ref{neq7}), 
\begin{equation}
\begin{split}
&\Delta \omega =\omega_1-\omega_2, \quad x=\nu_1-\nu_2=\Delta \omega t + \nu^{0}_1-\nu^{0}_2,\\ & x^0=x(t=0)=\nu^{0}_1-\nu^{0}_2. 
\label{neq8a}
\end{split}
\end{equation} 
From now on, it is implied that $k=1$ and $2$ are chosen in such a way that $\xi < 1$. It is also implied that $\arctan (y)$
is evaluated on the next branch  whenever  $|y|$ passes through $\infty$, and accordingly it takes the form
$\arctan (y)=p.v. \arctan (y)+n\pi$, where $p.v.$ indicates that the principal value is to be taken  and $n$ is a number of passages that  $|y|$ has undergone through $\infty,$
starting from $t=0$.

{It is useful to average $\dot \varpi_\mathrm{NI}$ over  the phase, $\nu_1-\nu_2$,  leading to the mean value defined through
 \begin{align}
 {\langle \dot \omega_\mathrm{NI}\rangle}={1\over 2\pi}
\int_{0}^{2\pi}dx \dot \varpi_\mathrm{NI},
\end{align}
From (\ref{nprec2}) it easily follows that
\begin{align}
{\langle \dot \omega_\mathrm{NI}\rangle}= - {JB_1\over S_1}.
\label{nprec2a}
\end{align}}

In order to find  $\Delta_k$, we separate it into two parts. Thus,  $\Delta_k=\Delta_k^{1}+\Delta_k^2$, where
$\Delta_k^{1}$ and $\Delta_k^2$ are solutions of (\ref{neq5}) with either the first or the second term on the right-hand side
being set to zero, respectively.
It is straightforward to find $\Delta^1_{k}$ in the form
\begin{equation}
\Delta^1_k={S_j\sin \delta^0_j \over S_k  \sin \delta^0_k  (\omega_k-\omega_j) J}T_{\perp,k}(\cos(x)-\cos (x^0)),
 \label{neq9}   
\end{equation}   
where we recall that in our approximation scheme  $T_{\perp,k}$ is  a function of $\delta_k^0$. 

Using eqns (\ref{neq7}), (\ref{neq8a}), and  (\ref{Tperp}), with the first term on r.h.s being neglected, we can rewrite (\ref{neq9}) in the form  $\Delta^1_k=A_{\perp,k}(\cos(x)-\cos (x^0))$,
where
\begin{equation}
A_{\perp, 1}={S_1\sin \delta^0_1\xi^2\over J(\tau_{\perp}-\xi)}, \quad A_{\perp,_2}=-{S_1\sin \delta^0_1\tau_{\perp}\over J(\tau_{\perp}-\xi)},
\label{neq9b}   
\end{equation}  


\noindent and
\begin{equation}
\tau_{\perp}={T_{\perp, 2}\over T_{\perp, 1}}\approx \left ({M_2\over M_1}\right)\left({\Omega_{r,2}\over \Omega_{r,1}}\right)^2
\left ({\sin (2\delta_2^0)\over \sin (2\delta_1^0)}\right).
\label{neq9c}   
\end{equation} 
We define a characteristic amplitude of variations of $\delta_k$ caused by the 'perpendicular'
torques, $A_{\perp}$, as
\begin{equation}
A_{\perp}=\sqrt{A_{\perp,1}^2+A_{\perp,2}^2}={S_1\sin \delta_1^0\over J|\tau_{\perp}-\xi|}\sqrt{\xi^4+ \tau_{\perp}^2}.
\label{neq9d}   
\end{equation}
From eq. (\ref{neq9d}) we see that that, when $S_1\sim S_2$, and given that each of the quantities  $\tau_{\perp},\xi,$ and $\sin (\delta_k^0))$
are expected to be  of order unity, ignoring the possibility that $\tau_{\perp}-\xi$ vanishes, a typical amplitude of variation due to the presence of perpendicular torques is an order of
$S/J$, where $S$ is either $S_1$ or $S_2$ 
\footnote{{The quantity $\tau_{\perp}-\xi$ vanishes when $\omega_1=\omega_2$,
and there is a potential divergence. However, this ignores the dependence of these frequencies on $\delta_k.$ 
We could then expect, assuming that $\Delta_1^1$ and $\Delta_2^1$ are comparable, that $\omega_1 -\omega_2 \propto  (\Delta_1^1)^{p},$
for some integer $p.$ This would lead to $\Delta_k^1$ being of characteristic magnitude $(S/J)^{1/(p+1)}.$ }}.  
%

When $\beta_k$ are set to be equal to $\delta_k^0$ in the expression (\ref{TX}) for $T_{\parallel,k}$, and 
the eccentricity is assumed to be a constant, it can be represented in the form $T_{\parallel,k}=-C_k\sin \delta_k \sin (2\hat \varpi_k)$,
where $C_k$ is a known positive constant depending on the orbital parameters and stellar properties.  Therefore, from eq. (\ref{neq5}) it
follows that 
$\Delta_k^2$ can be determined from the equation
\begin{equation}
{d\Delta^2_k\over dt}= {C_k \sin \delta_k \over S_k}\sin (2\hat \varpi_k),
\label{neq10a}   
\end{equation} 
which has an obvious formal quadrature,
 \begin{equation}
\Delta^2_k= {C_k \sin \delta_k \over S_k}\int_{0}^{t}dt^{'}  \sin (2\hat \varpi_k),
\label{neq10}   
\end{equation} 
where eqs (\ref{neq3}), (\ref{nprec0}), (\ref{neq7}), and (\ref{neq8}), and the assumption that the $\nu_k$ are linear functions of time can 
be used to determine the dependency of the integrand on time. 

The evaluation of eq. (\ref{neq10a}) can be simplified when  the average precession rate (\ref{nprec2a}) is used
in (\ref{nprec}) instead of $\dot \varpi_\mathrm{NI}$. 
In this case ${d\varpi / dt}\equiv \dot 
\varpi_\mathrm{tot}=\dot \varpi_{*}+{\langle \dot \omega_\mathrm{NI}\rangle}$ is constant, and, accordingly, $\varpi$ is a linear
function of time. We obtain from (\ref{nprec0}) and (\ref{neq10a})
\begin{equation}
\begin{split}
&{d\Delta^2_k\over dt}= {C_k \sin \delta_k \over S_k}\sin (2\hat \varpi_k)=\\&{C_k \sin \delta_k \over S_k}(\sin (2\varpi)\cos (2(\nu_k-\alpha))-\cos (2\varpi)\sin (2(\nu_k-\alpha))).
\end{split}
\label{neq10bn}
\end{equation} 
From eq. (\ref{neq3}), it follows that  
\begin{equation}
\begin{split}
&\cos (\nu_1-\alpha)=-{(1+\xi \cos (\nu_1-\nu_2))\over \sqrt {(1+\xi^2+2\xi \cos (\nu_1-\nu_2))}}, \quad \\
&\sin (\nu_1-\alpha)=-{\xi \sin (\nu_1-\nu_2)\over \sqrt {(1+\xi^2+2\xi \cos (\nu_1-\nu_2))}},
\end{split}
\label{neq10cn}   
\end{equation}  
while the expressions for $\cos (\nu_2-\alpha)$ and $\sin (\nu_2-\alpha)$ can be obtained from (\ref{neq10cn}) by
the substitution $\xi \rightarrow \xi^{-1}$. Thus, $\cos (2(\nu_k-\alpha))$ and  $\sin (2(\nu_k-\alpha))$ 
in (\ref{neq10bn}) are the functions of $x=\nu_1-\nu_2$ only. 
Note that the use of (\ref{nprec2a}) can be justified provided that a characteristic timescale 
of variations of $\nu_1-\nu_2$ in (\ref{nprec2}) is much shorter than the characteristic timescale
of the evolution due to the presence of 'parallel' torques,  which results in the condition
specified by the inequality (\ref{cr5}) below.

Equation (\ref{neq10bn}) can be easily solved in terms of Fourier series. We expand $\cos 2(\nu_1-\alpha)$ and $\sin 2(\nu_1-\alpha)$
as
\begin{equation}
\begin{split}
&\cos2 (\nu_k-\alpha)=c_{0,k}+\sum_{n=1}^{\infty}c_{n,k}\cos (nx), \quad \\
&\sin 2(\nu_k-\alpha)=\sum_{n=1}^{\infty}s_{n,k}\sin (nx),  
\end{split}
\label{neq1add}   
\end{equation} 
where we recall that $x=\nu_1-\nu_2=\Delta \omega t+x_{0}$, $x_{0}=\nu^{0}_1-\nu^{0}_2$ and use the fact that $\cos 2(\nu_1-\alpha)$ and $\sin 2 (\nu_1-\alpha)$
are even and odd functions of $x$, respectively. We then substitute (\ref{neq1add}) to (\ref{neq10bn}) to obtain
\begin{equation}
\begin{split}
{d\Delta^2_k\over dt}=&{C_k \sin \delta_k \over S_k}(c_{0}^{k}\sin (2\varpi)+ {}\\
&{1\over 2}\sum_{n=1}^{\infty}(c_{n,k}-s_{n,k})\sin (nx+2\varpi)-(c_n^{k}
+s_n^{k})\sin(nx-2\varpi)),
\end{split}
\label{neq2add}   
\end{equation} 
which has an obvious solution

\begin{equation}
\begin{split}
& \Delta^2_k=-{C_k \sin \delta_k \over S_k}\Biggl(c_{0,k}{(\cos (2\varpi)-\cos (2\varpi_{0}))\over 2\dot \varpi_\mathrm{tot}}+{}\\
&{1\over 2}\sum_{n=1}^{\infty}(c_{n,k}-s_{n,k}){(\cos (nx+2\varpi)-\cos (nx_{0}+2\varpi_{0}))\over (n\Delta \omega +2\dot \varpi_\mathrm{tot})}  \\
& -(c_{n,k} +s_{n,k}){(\cos (nx-2\varpi)-\cos (nx_0-2\varpi_0))\over  (n\Delta \omega -2\dot \varpi_\mathrm{tot})}\Biggr).
\label{neq3add}   
\end{split}
\end{equation}

When $|\Delta \omega | \ll |\dot \varpi_\mathrm{tot}|$
we can neglect variations
of $\gamma_k$ in time in  (\ref{neq10bn}), thus obtaining
\begin{equation}
\Delta^2_k\approx  -{C_k \sin \delta_k \over 2\dot \varpi_\mathrm{tot} S_k}(\cos (2\hat \varpi_k)-\cos (2\hat \varpi_{k,0})).
\label{neq10dn}   
\end{equation}

In the opposite limit  $|\Delta \omega | \gg \dot \varpi_\mathrm{tot}$, the terms in the summation series in (\ref{neq3add}) are expected 
to be much smaller than the term proportional to $c_{0}^{k}$, and we approximately have 
\begin{equation}
\Delta^2_k\approx -{c_{0,k} C_k \sin \delta_k \over S_k}{(\cos (2\varpi)-\cos (2\varpi_{0}))\over 2\dot \varpi_\mathrm{tot}}.
\label{neq4add}   
\end{equation}

Using the
expressions (\ref{neq10cn}) it is straightforward to obtain
\begin{equation}
c_{0,1}={1\over 2\pi}\int_{0}^{2\pi} d(\nu_1-\nu_2) \cos 2(\nu_1-\alpha)=1-\xi^2,
\label{neq10en}   
\end{equation}
when $\xi \le 1$ and $c_{0,1}=0$ otherwise, while the expression for $c_{0,2}$ is obtained from (\ref{neq10en}) 
by the substitution $\xi \rightarrow \xi^{-1}$. Recalling that we arrange the indices $1$ and $2$ in such a way that $\xi \le 1$ 
we find that only $\Delta^2_1$ is expected to vary
due to the presence of the parallel torque in the limit  $|\Delta \omega | \gg \dot \varpi_\mathrm{tot}$ and finally obtain 
\begin{equation}
\Delta^2_1\approx -{(1-\xi^2)C_1 \sin \delta_1 \over S_1}{(\cos (2\varpi)-\cos (2\varpi_{0}))\over 2\dot \varpi_\mathrm{tot}}.
\label{neq5add}   
\end{equation}

Comparing the asymptotic expressions (\ref{neq10dn}) and (\ref{neq5add}), we see that they differ only by a multiplicative
factor. In order to describe them in the same manner, we introduce a factor $f_k$ defined by the following rule. 
When $|\Delta \omega | \ll \dot \varpi_\mathrm{tot}$, $f_1=f_2=1$, while when $|\Delta \omega | \gg \dot \varpi_\mathrm{tot}$, 
$f_1=1-\xi^2$ and $f_2=0$. Making use of  this 
factor we obtain 
\begin{equation}
\Delta^2_k\approx -{f_kC_k \sin \delta_k \over S_k}{(\cos (2\varpi)-\cos (2\varpi_{0}))\over 2\dot \varpi_\mathrm{tot}},
\label{neq6add}   
\end{equation}
being valid in both limits.

\subsection{The effect of parallel torques}\label{parallelt}
Just as for variations due to the presence of the perpendicular torques, we can
estimate  a characteristic amplitude of variations due to the presence of the parallel torques that is
obtained from eq. (\ref{neq6add}) in the form
\begin{equation}
A_{\parallel}={1\over |\dot \varpi_\mathrm{tot}|}\sqrt{\left ({f_1^2C_1^2\sin^2(\delta_1)\over S_1^2}+{f_2^2C_2^2\sin^2(\delta_2)\over S_2^2}\right)}.
\label{neq10c}   
\end{equation}
From the condition $A_{\parallel} > A_{\perp}$, we can find a region in the parameter space of 
the problem, where the 'parallel' torques dominate over the 'perpendicular' torques in
the dynamical evolution. As we have discussed
earlier, the parallel torques are proportional to  $({\Omega_{k,r}/ \omega_{eq,k}})^2$, 
while perpendicular ones are proportional to  $({\Omega_{k,r}/ n_0})^2$, and, therefore, the latter
are expected to be much larger than the former in the weak tide limit where $n_{0}\ll  \omega_{eq,k}$.

On the other hand, from eq. (\ref{neq9d}), the amplitude  $A_{\perp}$ is proportional
to a small parameter $S_1/J$.
It is instructive to scale out these factors explicitly by redefining the amplitudes 
$\bar A_{\parallel}=({\omega_{eq,1}}/\Omega_{1,r})^2 A_{\parallel} $ and 
$\bar A_{\perp}= (J/S_1) A_{\perp}$. In this case the condition $A_{\parallel} > A_{\perp}$ is  expressed in the form
\begin{equation}
{J\over S_1}\left({\Omega_{r,1} \over \omega_{eq,1}}\right)^2 > {\bar A_{\perp}\over \bar A_{\parallel}}.
\label{neq10d}   
\end{equation}

Note that the r.h.s of (\ref{neq10d}) formally tends to infinity when parameters of the problem
are chosen in such a way that the condition
\begin{equation}
\dot \varpi_\mathrm{tot}=0,
\label{neq10e}   
\end{equation}
defining a so-called critical curve  (see  Section \ref{critical_curves} and also IP1), is satisfied. On the critical curve,
  $ A_{\parallel}$, being given by  (\ref{neq10c}),  and thus $\bar A_{\parallel}$
formally diverge, with eq. (\ref{neq10a}) accordingly becoming invalid. 
Accordingly, the simplified analysis presented in this Section is not valid near a critical curve, a situation that will be 
 discussed in Section \ref{critical_curves} below.

{Let us estimate what parameters of the system are expected to satisfy the inequality (\ref{neq10d}).
For this, we assume that both stars have approximately the same radii, masses, dimensionless moments
of inertia, and apsidal motion constants, $R$, $M$, $\tilde I$, and $k_2$, respectively, the frequency
$\omega_{eq,k}$ is represented as $\omega_{eq,k} = \Omega_* \tilde{\omega}_{eq,k}$, where $\Omega_{*}=\sqrt{{GM/ R^3}}$, and we assume
that $\Omega_{r,2}\sim\Omega_{r,1} \equiv \Omega_{r},$  and that $\bar A_{\perp}=1$.\footnote{ But note that when  $|\Delta \omega | \gg \dot \varpi_\mathrm{tot}$ we assume that $\xi$ is not too close to one.} Additionally, all trigonometric factors in 
all expressions of interest are set to unity.
The condition (\ref{neq10d}) can then be written as
{\bf \begin{align}
 f_1 C_1J>S^2|\dot \varpi_\mathrm{tot}|.\label{cond60a}
 \end{align}}

From eq. (\ref{TX}) we find, adopting the above simplifications, that
{\bf \begin{align}
C_1=\frac{9f_1k_2MR^2}{4}\frac{e^2(1+e^2/6)}{(1-e^2)^{9/2}}(2\beta_*+1) n_0^2 \left(\frac{\Omega_r}{\tilde{\omega}_{eq,k} \Omega_*}\right)^2 \tilde a^{-3},
\end{align}}
where $\tilde a = a/R.$
 In line with the above assumptions, we set \\
$J\approx L\sim \frac{1}{2}MR^2 \tilde a^2 n_0\sqrt{(1-e^2)}$. 
The condition (\ref{cond60a}) then yields 
{\bf \begin{align}
\frac{9f_1k_2}{8}\frac{e^2(1+e^2/6)}{(1-e^2)^{4}}(2\beta_*+1) n_0 \left(\frac{n_0}{\tilde{\omega}_{eq,k} \Omega_*}\right)^2 \tilde a^{-1}> {\tilde I}^2|\dot \varpi_\mathrm{tot}|.\label{61}
\end{align}}
 
It is also assumed that $\dot \varpi_\mathrm{tot} \sim \dot \varpi_\mathrm{NI}$.  Accordingly, we introduce
$\dot {\tilde \varpi}_\mathrm{tot}= \dot \varpi_\mathrm{tot}/ \dot \varpi_\mathrm{NI}$ and estimate  $\dot \varpi_\mathrm{NI}$
as $|\langle \dot \varpi_\mathrm{NI}\rangle| \sim {\cos{\delta} k_2 \Omega_r {\tilde a}^{-3}/ (\tilde I (1-e^2)^{3/2})}$, 
where we have used (\ref{nprec2a})
and (\ref{b2})-(\ref{b3}). Making use of this in (\ref{61}), we obtain
{\bf \begin{align}
\frac{9f_1}{4}\frac{e^2(1+e^2/6)}{(1-e^2)^{5/2}}(2\beta_*+1) \sigma^{-1} \tilde a^{-1} \tilde{\omega}^{-2}_{eq,k} \cos^{-1}{\delta}> {\tilde I}|\dot {\tilde \varpi}_\mathrm{tot}|,
\label{61n}
\end{align}}
where $\sigma=\Omega_r/n_0$.}

When $\sigma $ is  large enough for $\dot \varpi_\mathrm{NI}$ to dominate the apsidal precession
 (but not so large that the spin angular momentum  becomes comparable to the orbital
angular momentum), we expect  $\dot {\tilde \varpi}_\mathrm{tot}\approx 1$. In this regime, the inequality  (\ref{61n}) is expected to be valid for sufficiently
small values of  $\tilde a$ and  $\sigma$, together with sufficiently large $e$. In addition, it is clear that the inequality (\ref{61n}) will be satisfied
sufficiently close to critical curves, where $\dot {\tilde \varpi}_\mathrm{tot}=0$.

{\subsection{The evolution near a critical curve}
\label{critical_curves}

Let us now assume that the parameters of the problem are such that $\dot \varpi_\mathrm{tot}$ is small enough for eq. (\ref{neq10e}) to be approximately valid. We also make the additional assumption that we can separate the time evolution of quantities
averaged over a 'short' timescale  $\sim {\rm max} ({S_k/ T_{\perp,k}})$ that characterises the oscillations of the second term on the right-hand side of (\ref{neq5}) (see  (\ref{Tperp})). 
 Similarly, we average  $\dot \varpi_\mathrm{NI}$ over the 'short' timescale and use eq. (\ref{nprec2a}) to evaluate this average. \footnote{  In doing this we assume that the angle $\nu_1-\nu_2$ changes linearly with time
so that the time average may be performed  by  taking the average for the angle   varying  uniformly over the interval $(0,2\pi).$ }
The evolution equations for the averaged inclination angles, $\delta_k,$  that are to be obtained,  do not depend on $ T_{\perp,k}$.
In general, having obtained the secular evolution, the effect of the neglected fast variations may be subsequently considered. Nevertheless, here we restrict ourselves to a qualitative sketch. 

Following the  discussion leading to the finding that $\Delta^{2}_2,$ 
 corresponding to the component which has the smaller
  component of spin angular momentum perpendicular to the total angular momentum, 
undergoes only relatively low amplitude
 variations  
  in  Section \ref{approxdelta},
   it follows that only $\Delta_1$ is expected
to evolve significantly under the assumptions adopted here. 
 Thus we consider the evolution of only this quantity, setting $\Delta=\Delta_1$ below.  

In order to calculate the time-averaged $\dot \varpi$, we take into account the next order correction in the expansion of (\ref{nprec2a}) in
$\Delta$. Noting that $\langle \dot \varpi\rangle  \propto \cos\delta_1 $,  we obtain 
\begin{align}
\langle\dot \varpi\rangle \approx \langle \dot {\varpi}_\mathrm{tot}^{0}\rangle  -\langle \dot \omega^{0}_\mathrm{NI}\rangle\tan (\delta_1^{0})\Delta,
\label{cr1}
\end{align}
where the index $(0)$ means that the corresponding quantities are evaluated with  $\delta_k = \delta_k^0$. 
If needed, the full value of  $\dot \varpi$ can be found by adding the rapidly varying contribution ${\dot\varpi}_\mathrm{fast}.$
Thus $ \dot \varpi=  \langle\dot \varpi \rangle+  \dot \varpi_\mathrm{fast}.$

Writing the first term on right-hand side  of (\ref{neq5}) in the form of 
the right-hand side of  equation (\ref{neq10bn}),  and  noting  that $c_{0,1} = 1-\xi^2$,  equation (\ref{neq5}) takes the form
\begin{align}
{d\Delta\over dt}=(1-\xi^2){C_1\over S_1}\sin \delta^{0}_{1}\sin (2\varpi ) +{\cal K}_k^0,
\label{cr2}
\end{align}
where, recalling (\ref{neq1add}),  ${\cal K}_k^0$ is the sum of the terms with factors of the form $\cos nx,$ or $\sin nx$ with $n\ge 1$ that occur in the
 series in (\ref{neq10bn}) together with the second term on the right-hand side of (\ref{neq5}), which is assumed to be rapidly varying.
 
To proceed, we differentiate (\ref{cr1}) with respect to time and then
substitute  $d\Delta/dt$ using  (\ref{cr2}). In this way, we obtain an equation  governing a standard pendulum, but with the addition of high-frequency forcing (see also IP1)
\begin{align}
{d^2 \theta \over dt^2}= -\Omega_{\parallel}^2\sin \theta +\frac{d\dot \varpi_\mathrm{fast}}{dt} - {\cal K}_k^0\tan\delta_1^0 \langle \dot \omega^{0}_\mathrm{NI}\rangle,
\label{cr3}
\end{align}
where $\theta=2\varpi$ and 
{\bf \begin{align}
\Omega_{\parallel}=\sqrt{{2J(1-\xi^2)C_1|\Omega_{Q,1}|\sin^2\delta_{1}^{0}\over S_1^2}}.
\label{cr4}
\end{align}}
The characteristic time $\Omega_{\parallel}^{-1}$ provides the timescale of 'slow' evolution due to the presence of 
the parallel torque, and, accordingly,  in order to separate the secular changes from those occurring on the more rapid timescale,
we should have (see eq. (\ref{neq8a}))
\begin{align}
|\Delta \omega| \gg \Omega_{\parallel}.
\label{cr5}
\end{align}

Eq. (\ref{cr3}), with the last two high-frequency forcing terms on the right-hand side neglected,
 has a well-known general solution in terms of incomplete elliptic integrals. We do not provide it here, limiting ourselves to
a qualitative analysis. It is easy to see that, in these circumstances, (\ref{cr3}) possesses a first integral of the form
\begin{align}
E={{\dot \theta}^2\over 2}+\Omega_{\parallel}^2(1-\cos \theta).
\label{cr6}
\end{align}
When $t=0$,  $\Delta=0$ by definition, and, accordingly, $\dot \theta (t=0)=2\dot \varpi_\mathrm{tot}$, while $\theta=2\varpi_\mathrm{in}$, where $\varpi_\mathrm{in}=\varpi (t=0)$.
Therefore, $E = 2({\dot \varpi_\mathrm{tot}}^2+\Omega^{2}_{\parallel}\sin^2(\varpi_\mathrm{in}))$.

When $E < E_\mathrm{crit}=2\Omega^{2}_{\parallel}$ the motion of $\theta $ is periodic and we have coupled librations of  $\Delta$ and $\varpi$ with the period
\begin{align}
P_{\parallel}=4K({\bf k})\Omega_{\parallel}^{-1}, \quad {\bf k}=\sqrt{{E\over 2\Omega_{*}^2}},
\label{cr6a}
\end{align}
where $K(k)$ is the complete elliptic integral of the first kind. This regime is analogous to that discussed in IP2. 
The condition $E=E_\mathrm{crit}$ defines the separatrix solution dividing
the regimes of librating and circulating $\varpi$. 
When $E \gg E_\mathrm{crit}$ the solution (\ref{neq5add}) is reproduced provided that
the condition
\begin{align}
\dot \varpi_\mathrm{tot} \gg \Omega_{\parallel}
\label{cr6b}
\end{align}
is satisfied. 

In order to find variations of {\bf $\Delta$} for the separatrix solution, we first substitute $E=2\Omega_{\parallel}$ in (\ref{cr6}), which leads to
\begin{align}
\dot \varpi =\Omega_{\parallel}\cos (\varpi).\label{xx1}
\end{align}
From (\ref{xx1}), we see that $\varpi$ moves between $-\pi/2$ and $\pi/2$, taking an infinite time. To find $\Delta$,
 we use (\ref{xx1}) to change the independent variable in (\ref{cr2}) from time to $\varpi$. We then integrate the result
over $\varpi$ from $0$ to $\pi/2$ to obtain the maximal amplitude 
{\bf \begin{align}
\Delta^\mathrm{max}={2(1-\xi^2)C_{1}\sin (\delta_{1}^{0})\over S_{1}\Omega_{\parallel}}=\sqrt{{2(1-\xi^2)C_1\over J |\Omega_{Q,1}|}}, 
\label{cr8}
\end{align}}
where we use (\ref{cr4}) to obtain the last equality.

At this point, we remark that the effect of the last two terms on the right-hand side of (\ref{cr3}), 
that are expected to produce high-frequency forcing of the pendulum, need to be considered.
The effect of these is well known in studies of dynamical systems and celestial mechanics
(see e.g.  Rathe et al. 2022). A stochastic layer is produced in the region of the separatrix, with motions
chaotically changing between libration and circulation. The size of this depends on the 
characteristics of the forcing. Additionally, considering high-frequency forcing reduces the maximum amplitude for long-term librating solutions, $\Delta^\mathrm{max}$. 
}

\section{Numerical modelling
}
\subsection{The significant timescales governing orbital evolution}

Before describing the results of our numerical simulations we find it convenient to express significant characteristic timescales in forms that we find to be useful.

\subsubsection{The timescale associated with the apsidal precession rate}
The first important timescale, $t_\mathrm{aps}$,  characterises the apsidal precession rate due to the tides raised in the primary component for a
fiducial orbit with zero eccentricity. To estimate this timescale we make use of
 Eq. (\ref{Apse}), in the limiting case of  a circular orbit, noting  that the equivalent result  can also be inferred from IP1 (see their Eq. (20)), with the consequence that
\begin{align}
t_\mathrm{aps} = 4.2 \times 10^{5} \; \mathrm{yr}\; q^{-1} (1+q)^{-1/2}&\left(\frac{M_1}{5.1\: \mathrm{M_{\odot}}}\right)^{-1/2}\left(\frac{R_1}{2.79 \:\mathrm{R_{\odot}}}\right)^{-5} \nonumber\\&\left(\frac{a}{0.20 \:\mathrm{AU}}\right)^{13/2}\left(\frac{k_{2,1}}{8.5 \times 10^{-4}}\right)^{-1},
\label{tscl1}
\end{align}
with $q = M_2/M_1$.
Note that, in (\ref{tscl1}) and similar expressions below, we have scaled the parameters defining the system using the values characterising the DI Her system. 
\subsubsection{The timescale associated with evolution driven by the action of the parallel component of the torque  $T_{\parallel,1}$ due to the primary in the neighbourhood of critical curves}
The second timescale of interest is $t_\mathrm{\parallel} = 2\pi / \Omega_{\parallel}$,  with $ \Omega_{\parallel}$ being given by  Eq.(\ref{cr4}),
which characterises the evolution governed by the parallel component of the torque exerted by the primary component, $T_{\parallel,1}$.
In particular, it gives the timescale associated with librations in the neighbourhood of critical curves.
 This timescale is estimated as 
\begin{multline}
t_\mathrm{\parallel} = 7.6 \times 10^{4}\; \mathrm{yr}\; q^{-3/2} D_1 E_1 \left(\frac{\sin\delta_1}{0.96}\right)^{-2} \left(\frac{\tilde I_1}{8.6 \times 10^{-2}}\right)\left(\frac{M_1}{5.1\: \mathrm{M_{\odot}}}\right)^{-1/2}\\\left(\frac{R_1}{2.79 \:\mathrm{R_{\odot}}}\right)^{-3} 
\left(\frac{a}{0.20 \:\mathrm{AU}}\right)^{9/2}\left(\frac{k_{2,1}}{8.5 \times 10^{-4}}\right)^{-1}\left(\frac{\Omega_{r,1}}{6.6 \times 10^{-5}\: \mathrm{rad/s}}\right)^{-1}\\\left(\frac{\Omega_\mathrm{eq,1}}{2.1 \times 10^{-3}\: \mathrm{rad/s}}\right),
\label{tscl2}
\end{multline}
with $D_1 = \left((2 \beta_{*,1} +1)/2.41\right)^{-1/2}$ \\and $E_1 = (e^{-1}(1-e^2)^{3}(1+e^2/6)^{-1/2})/0.78$.
\subsubsection{The timescale for the evolution of $\nu_1 -\nu_2$}

Finally, the third potentially significant  timescale, 
expressed as  $t_\mathrm{\perp} = 2\pi / |\Delta \omega|$, is appropriate to the variation of $\nu_1-\nu_2$ occurring as a result of the action of
 $T_{\perp}$. This timescale
 can be estimated with the help of eqns (\ref{neq2}) and (\ref{neq7}) and is readily  expressed as
\begin{multline}
t_\mathrm{\perp} = 1.8 \times 10^{3}\; \mathrm{yr}\;q^{-1}E_2\left(\frac{\cos\delta_1}{0.27}\right)^{-1}\left(\frac{\tilde I_1}{8.6 \times 10^{-2}}\right)\left(\frac{R_1}{2.79 \:\mathrm{R_{\odot}}}\right)^{-3}\\\left(\frac{a}{0.20 \:\mathrm{AU}}\right)^{5}\left(\frac{k_{2,1}}{8.5 \times 10^{-4}}\right)^{-1} \left(\frac{\Omega_{r,1}}{6.6 \times 10^{-5}\: \mathrm{rad/s}}\right)^{-1}  
\left(1 - \frac{F_2}{F_1}\right)^{-1},
\label{tscl3}
\end{multline}
where $E_2 = \left((1-e^2)/{0.74}\right)^{3/2}$ and $F_\mathrm{k} = k_\mathrm{2,k} \Omega_\mathrm{r,k} R_\mathrm{k}^3/{(M_\mathrm{k}^2 \tilde I_\mathrm{k})}$.
However, note that, from the discussion of Section \ref{aqual}, the occurrence of oscillations on this timescale requires that both $S_1$ and $S_2$ are not zero.
In addition, we remark that (\ref{tscl3}) is
 valid only as long as the first term on the left-hand side of (\ref{Tperp}) is small compared to the second one.

One can see that, for parameters appropriate for DI Her, eqns (\ref{tscl1})--(\ref{tscl3}) lead to the expectation that, as long as we are not dealing with near-polar orbits {and $1 - F_2/F_1$ is not close to zero},  $t_\mathrm{\perp} < t_\mathrm{\parallel} < t_\mathrm{aps}$. These inequalities also hold for DI Her. On the other hand, for orbits very close to polar with $\cos\delta_1 \ll 1$,  eqns (\ref{tscl1})--(\ref{tscl3}) imply that we can have $t_\mathrm{\perp} > t_\mathrm{\parallel}.$
Also, note that $t_\mathrm{aps}$ can be comparable with $t_\mathrm{\parallel}$ for orbits with a significant eccentricity.

\subsection{Numerical treatment of the tidal evolution}\label{Numevol}
Although our discussion can be made quite general, in this Section, we focus on systems with parameters similar to those of the DI Her system. In order to discuss this in a quantitative manner, we require stellar models for the components.
\subsubsection{Models for DI Her}
In Appendix \ref{appendixA}, we outline the approach we used to calculate
stellar models of the components of the 
 DI Her system, and, in Appendix \ref{sec:method}, we explain how we apply those models to determine the parameters required to quantitatively specify the expected tidal evolution 
 of the system.
 Hereafter, the adopted stellar parameters correspond to the models \texttt{M51PY} and \texttt{M44PY} given in Table~\ref{table1} for the primary and secondary components, respectively. 
 We note that selecting other pairs of models listed in Table~\ref{table1} produces only minor effects on our results.

\subsection{The critical curves}\label{critc}
We begin our numerical treatment of DI Her
with an analysis of the critical curves.  Here, these are defined by setting the time average of $d\varpi/dt$, as given by eq. (\ref{nprec}), 
to zero, thus $\varpi_* +\langle\dot \varpi_\mathrm{NI}\rangle=0,$
with $\langle \dot \varpi_\mathrm{NI}\rangle$ given by eq. (\ref{nprec2a}). 
The time average is taken to remove high-frequency oscillations occurring on the shortest characteristic time indicated above, namely  $t_{\perp}.$ 
Note that this was not done in IP2 because, when $S_2=0,$ such high-frequency oscillations are not present.
{ Note also that, after the time averaging process, for small $S_k/J,$  the distance to a critical curve does not explicitly depend on the orbital inclination, $i,$ 
but on $\sigma$, $\tilde a$, $e,$ and $\delta_k.$
We note the constraint that the periastron radius should be larger than the sum of the radii of two components. In addition,  rotational angular velocities should lie below the break-up limit. 
These  two restrictions lead to  $\tilde a > (R_1 + R_2)/(R_1(1-e))  \approx 2/(1-e)$ and $\sigma =\Omega_r/n_0 < \sqrt{\tilde a^3 /(1+q) } \approx \tilde a^{3/2}/\sqrt{2}$, which are imposed below.

In this Section, for simplicity, we assume $\delta_1 = \delta_2 \equiv \delta$. Additionally, we assume that both components rotate with
angular velocity $\Omega_r,$  with $ \sigma=\Omega_r/n_0$.  The masses and radii of the components are $M_1 = 5.1 \; M_{\odot}$, $M_2 = 4.4 \; M_{\odot}$, $R_1 = 2.79 \; R_{\odot}$, $R_2 = 2.59 \; R_{\odot}$, in accordance with our \texttt{MESA} models.

In Fig.~\ref{fig1_C}, we show the critical curves on $(\tilde a$, $\sigma)$ plane obtained with different values of $e$ and $\delta$, specified in radians. As shown later in this Section, the system cannot be on the critical curve unless $\delta > \pi/2.$ Thus only the systems with retrograde rotation are considered here. The overall shape of the critical curves is in agreement with the results presented for the retrograde case in Fig. 3 of IP1. Notably, there are two branches of a critical curve, giving two values of ${\tilde a}$ for a specified value of $\delta,$ which is expected when orbital angular momentum and spin vectors are close to anti-alignment. With decreasing $e$ and increasing $\delta$, the critical curves reach smaller $\sigma$ values.

Additionally, we calculate the rate of precession of the longitude of periapsis, $d\Pi/dt$, for a system located on the critical curve. This quantity is useful from the observational point of view as it normally represents measured apsidal advance rates \citep[see e.g.][]{claret2010}. It is defined as the sum of the precessional rate of the line of nodes, $d \Omega /dt$, and the precessional
rate of the line of apsides with respect to the line of nodes, $d \varpi /dt$. 

In the limit of small $S_k/J$ (i.e., for small $i$), in which we are working, there is no dependence on $\delta.$ 
For fixed $\sigma,$ $M,$ and $R,$ 
$d\Pi/dt$ depends only on ${\tilde a}$ and $e.$   
This can be explicitly demonstrated from  (\ref{NI0}), if we  
 replace $\cot(i)$ by  $(\cos{i}-1)/\sin{i}$  and use the fact
that $\dot \varpi_\mathrm{NI}=-\omega_{*}$ on a critical curve in this approximation\footnote {{ We remark that, as indicated in Section \ref{aqual}, 
 ${\dot \varpi}_\mathrm{R}$ contributes to ${\dot \varpi}_*$ along with $\dot \varpi_\mathrm{T}$ and $\dot \varpi_\mathrm{E}$. However, we find that 
neglecting this contribution makes only very small changes to the particular numerical results presented here.}}. We find that 
$d\Pi/dt={\omega_{*}/ \cos (i)}\approx \omega_{*}$, where we recall that $i$ is small for these systems.
Our results for $d\Pi/dt$ are illustrated in Fig.~\ref{fig2_C}. One can see that this precessional rate monotonically decreases with the normalised orbital separation. There are two pronounced slopes representing different branches in Fig.~\ref{fig1_C}. The transition between these two branches marks the location in the parameter space where $\dot \varpi_\mathrm{E}$ and $\dot \varpi_\mathrm{T}$ become comparable. It is clear that increasing eccentricity leads to growth in $d\Pi/dt$. 

Nonetheless, the value of $\delta$ is essential for the determination of the accessibility of a critical state for a given system, 
as we demonstrate in Fig.~\ref{fig3_C}. Here, we show the critical curves in the $(\tilde a, \delta)$ plane for different rotation rates and eccentricities. 
The most important feature is that, below a certain $\delta$ value, depending on $\sigma$ and $e$, a critical state is no longer possible. 
Large rotation and eccentricity also restrict the possibility of the emergence of a critical state. 
In particular, there are no critical curves for $e = 0.9$ in the two lower panels that represent cases with rapid rotation.
 Another significant detail is the disappearance of the second branch for slow rotation and small eccentricity. 

\subsubsection{ Critical curves and DI Her}\label{critDIHer}

{ According to \citet{Philippov}, the spin inclinations are given by  
$\delta_1 = 62^\circ \pm 17^\circ$,
 and $\delta_2= 80^\circ \pm 20^\circ$, which would easily allow for the secondary to be retrograde and
 the primary also to be retrograde at the $2.2\sigma$ level.
  More recently,  \cite{Liang} estimated the  spin inclinations of the components of DI Her to  be $\delta_1 = 75^\circ \pm 3^\circ$,
 and $\delta_2= 80^\circ \pm 3^\circ$,
indicating that both are less than $\pi/2$. In this case, DI Her is unlikely to be close to a critical curve 
 if $d\Pi/dt >0.$ 
We note that, from the discussion in Section  \ref{critc} (see also eqns (\ref{nprec}) and (\ref{nprec2a})), it is possible to achieve a critical state with only one component being retrograde. Indeed, the secondary component appears close to being retrograde raising the possibility of libration of $\delta_2$ about a value slightly greater than $\pi/2.$
We also note that, to obtain their results, both \cite{Philippov} and \cite{Liang} made fits to the observations employing a dynamical model ignoring the torques of the form $T_{\parallel,k}$. This can have significant effects (see Section \ref{caseB} below) and may alter error estimates.}

 In Figs.~\ref{fig1_C}--\ref{fig3_C}, we mark the location of DI Her with a dark cross {in the panels that illustrate planes that contain the parameters
  suggested for this system, namely the top left panel in Fig. ~\ref{fig1_C} and bottom right panel in Fig. ~\ref{fig3_C}.} We recall that, for DI Her, $e \approx 0.51$, $\sigma_1 \approx 9.6$, and $\sigma_2 \approx 10.9$, see \cite{Liang}. According to our estimates, DI Her {may be brought to a state} close to critical curves in the $(\tilde a,\sigma)$ plane if one slightly increases $\delta_k$ to make the rotation
of the components retrograde.

 We show the position of the system with the   parameters determined for DI Her, but with
$\delta=\delta_1=\delta_2=1.7$ in the top left panel of Fig. \ref{fig1_C}. It is seen from this Fig. that, in order to bring the system precisely on the critical curve, one should reduce the rotation rate of its components by a factor of $\approx 2$.

\begin{figure*}
\begin{multicols}{2}
    \includegraphics[width=\linewidth]{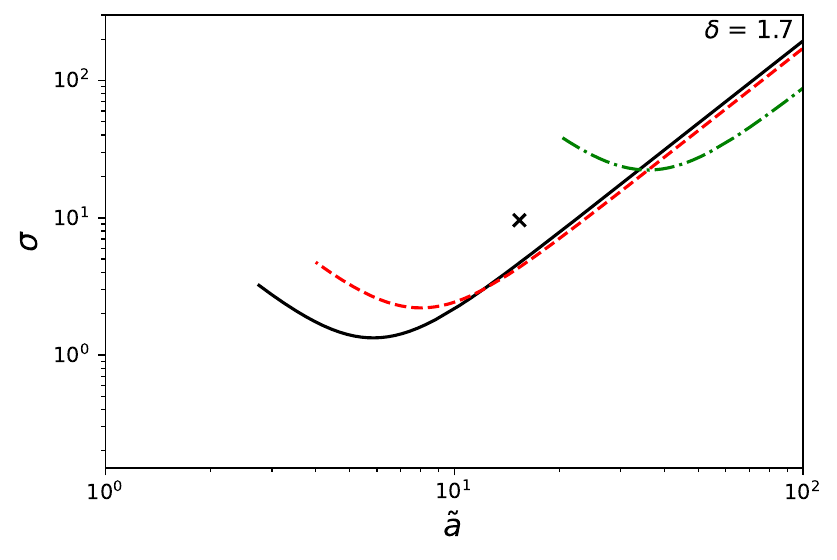}\par 
    \includegraphics[width=\linewidth]{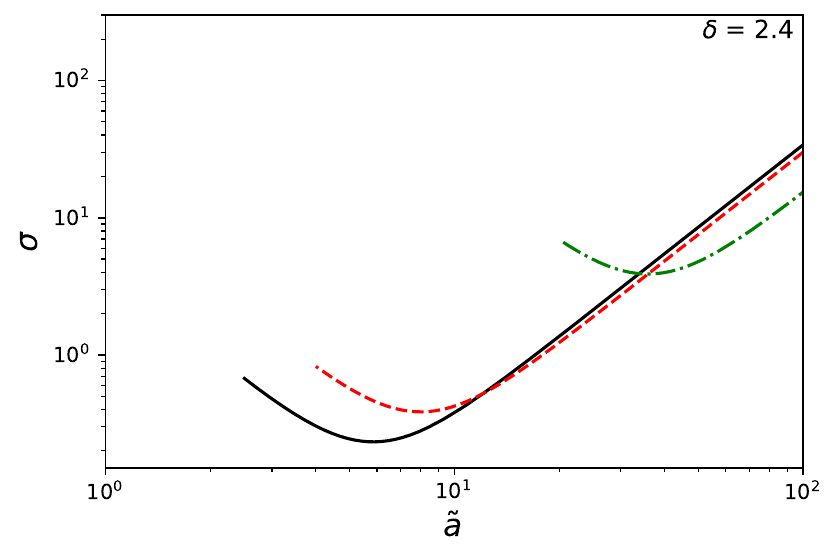}\par 
    \includegraphics[width=\linewidth]{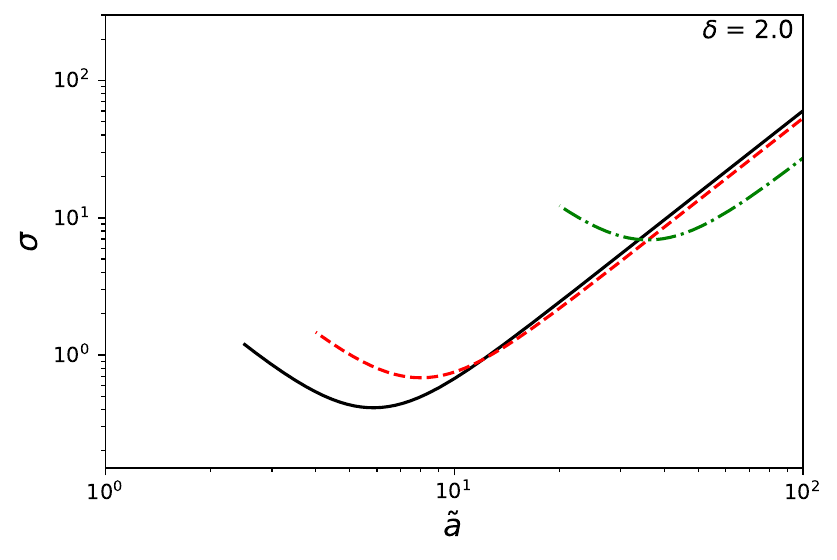}\par 
    \includegraphics[width=\linewidth]{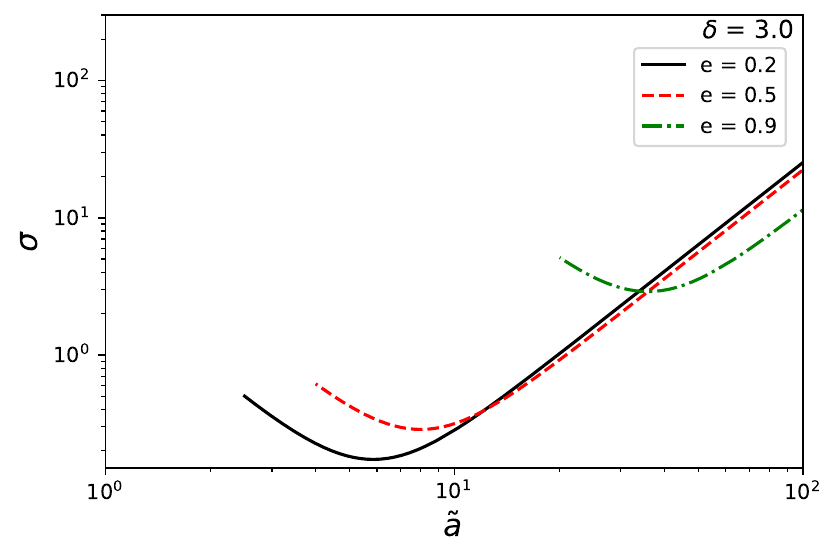}\par
    \end{multicols}
\caption{Critical curves in the  $(\tilde a, \sigma)$ plane. Dark solid, red dashed, and green dash-dotted lines correspond to $e = 0.2$, $0.5$, and $0.9$, respectively. Each panel represents a different $\delta$ value specified in the top right corner. Note that, in order to be on a critical curve {for the values of $\tilde a$
and $\sigma$ considered}, it is necessary that  $\delta > \pi/2.$
The actual location of DI Her is depicted by the dark cross in the top left panel.}
\label{fig1_C}
\end{figure*}

\begin{figure*}
\centering
    \includegraphics[width=100mm]{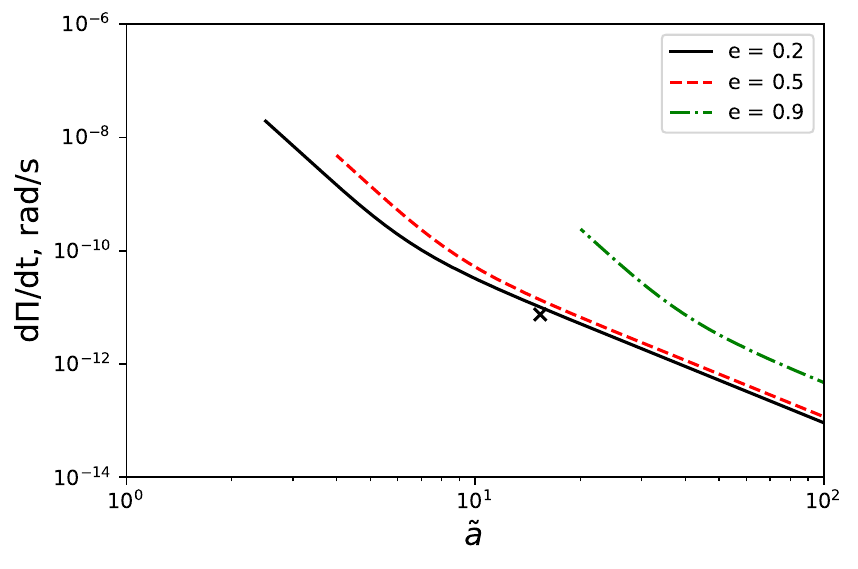}\par 
\caption{The rate of precession of the longitude of periapsis for a system on the critical curves. The parameters and designations are the same as in Fig.~\ref{fig1_C}.}
\label{fig2_C}
\end{figure*}

\begin{figure*}
\begin{multicols}{2}
    \includegraphics[width=\linewidth]{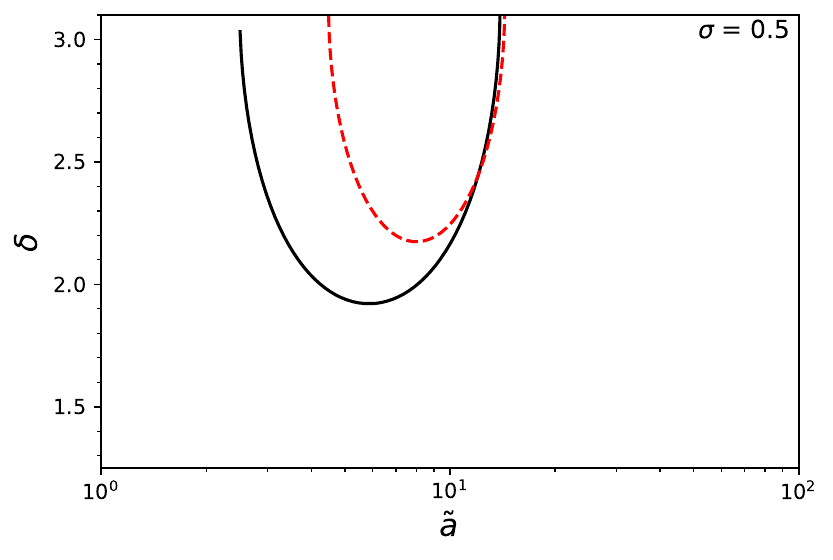}\par 
    \includegraphics[width=\linewidth]{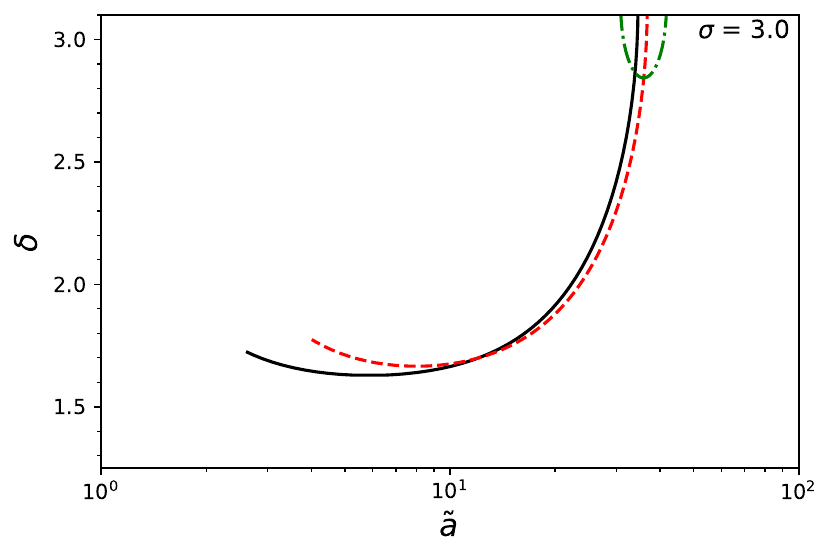}\par 
    \includegraphics[width=\linewidth]{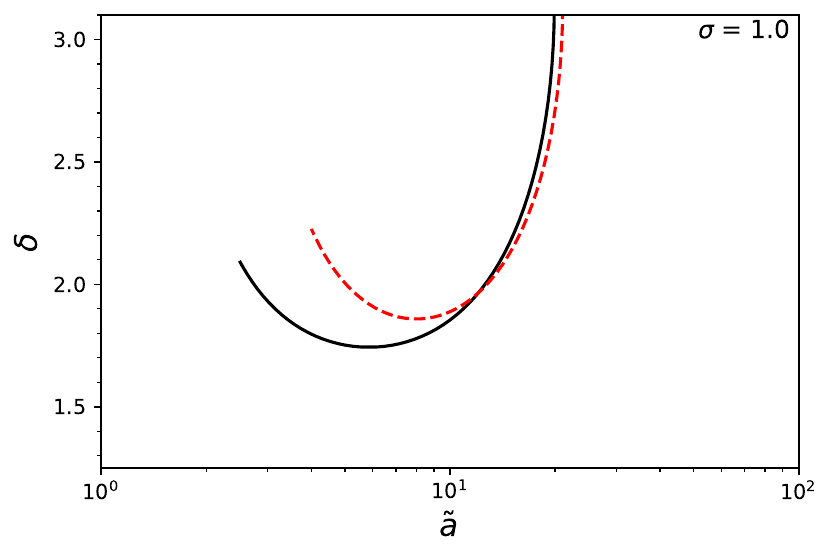}\par 

    \includegraphics[width=\linewidth]{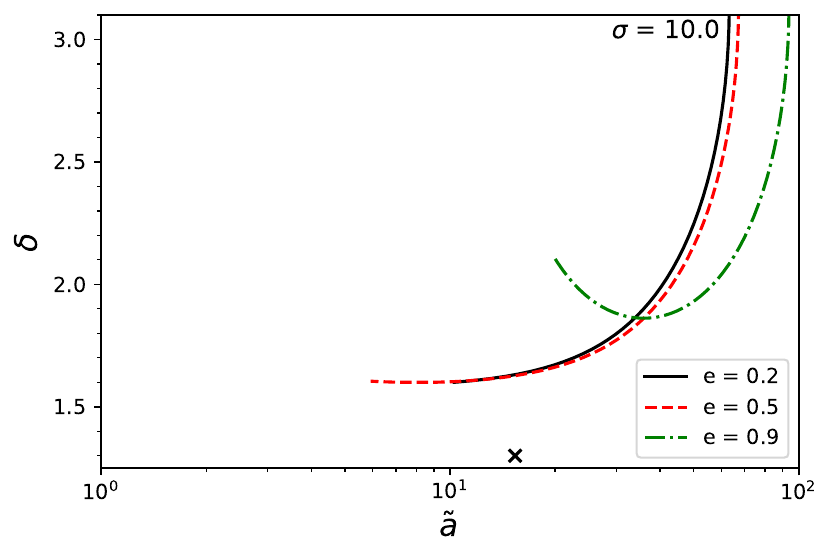}\par
    \end{multicols}
\caption{Critical curves in the $(\tilde a, \delta)$ plane. Each panel represents a different $\sigma$ value specified in the top right corner.}
\label{fig3_C}
\end{figure*}
\subsection{The orbital evolution in the vicinity of a critical curve}\label{orbitalcritc}
In this Section, we discuss the results of numerical simulations of the orbital evolution of binary systems undergoing tidal interactions of the kind described above.
To do this we make use of the Python \texttt{SciPy} routine \texttt{odeint}. Using the function \texttt{odeint}, we solve the ordinary differential equations (eqns ~(\ref{eq14}) -- (\ref{eq18}) and eq.~(\ref{prec})) that govern the evolution of the system as an initial value problem.  Accordingly, initial orbital parameters need to be specified.
\subsubsection{Specifying the initial parameters for orbital evolution}

The fixed quantities characterising a system are $a$ and  $\sigma,$ 
while initial values need to be provided for, $e, $  $\delta_k$ and $\nu_k,$ for  $k=1,2,$  
 and the orbital inclination, $i$ (see Section \ref{finaleq}).  There is a 
range of acceptable values for the latter that depends on the other parameters specified. This acceptable range is
[$i_\mathrm{min}$, $i_\mathrm{max}$], where both $i_\mathrm{min}$ and $i_\mathrm{max}$ can be inferred from eqns (\ref{j3}) and (\ref{neq4}). 
The maximum inclination angle $i_\mathrm{max}$  corresponds to the case where $\alpha - \nu_1 = \alpha - \nu_2 = \pi$ (i.e., the spin projections on the plane containing $({\textbf e}_x, {\textbf e}_y)$ are anti-aligned with the projection of the orbital angular momentum). 
From eq. (\ref{neq4}), which applies in the limit of small $S_k/J,$ $i_\mathrm{max}= (S_1\sin\delta_1+S_2\sin\delta_2)/J.$  
The minimum inclination, $i_\mathrm{min}$, 
corresponds to  the case where $\alpha - \nu_1 = \pi$ and $\alpha - \nu_2 = 0$, given that $S_1 > S_2$. (i.e., the projection of a smaller (larger) spin is aligned (anti-aligned) with the projection of the orbital angular momentum). From eq. (\ref{neq4}),  $i_\mathrm{min}= |S_1\sin\delta_1-S_2\sin\delta_2|/J.$  
To make sure that our initial orbital configuration is physically possible, we set $i_0 = (i_\mathrm{min} + i_\mathrm{max})/2$. We remark that
as the explored variables span a few orders of magnitude, the ranges of acceptable inclinations corresponding to different setups may not overlap.  

In this Section, we consider two representative sets of values of the conserved quantities and the initial values of the orbital parameters. These are denoted as case A and case B. 
Case A is used to illustrate the 
theory developed in Section \ref{critical_curves}. Accordingly, the  parameters associated with the orbit and  the angular velocities 
 of the  stellar components as well as the inclination angles
between the stellar spins and the conserved total angular momentum are chosen 
in such a way that that both librating and circulating regimes of the evolution of the apsidal angle can occur.

Case B illustrates a system with parameters chosen to be close to those of DI Her.
 However, as we have discussed above the parameters appropriate to  DI Her locate the system away from,  but relatively close to a critical curve. 
 In order to reach the proximity of a critical state, we chose the initial values of the inclinations, $\delta_k, $ to be very close to  $\pi/2$. In this case, it turns out that $t_{\parallel} < t_{\perp}$.    This is counter to  the assumption made in Section \ref{critical_curves}
   when describing motion close to a critical curve. This assumption leads to libration or circulation of the apsidal angle that undergoes small high-frequency perturbations. Accordingly,  case B illustrates a different regime of evolution, where the high-frequency terms are not necessarily small.
The details of the specification of the parameters corresponding to these cases and the descriptions of their dynamical evolution are given below. 

\subsubsection{Case  A}

Case A is taken to be a highly-eccentric system ($e = 0.9$), composed of two stars with equal rotation periods of $P_\mathrm{r,k} = 10.7$ days. The semi-major axis $a = 0.44$ AU. The initial value of $\varpi$ is set to zero. The dimensionless moments of inertia of two stars are fixed at $\tilde I_1 = 8.6 \times 10^{-2}$ and $\tilde I_2 = 4.3 \times 10^{-2}$, thus we decrease the moment of inertia of the secondary as compared to the primary  in order 
to make the ratio $S_1 B_2/(S_2 B_1)$ (see Sections \ref{aqual} and  \ref{approxdelta})  close to 2.
 In order to bring the system to the critical curve, we vary  $\delta,$  the initial value of both $\delta_1$ and $\delta_2,$
 while keeping the remaining stellar parameters (e.g., stellar mass and radii, overlap integrals, and equilibrium frequencies) the same as those for DI Her. The value of the parameter $\xi$ is approximately equal to $0.37$.
 
The left panel of Fig.~\ref{fig4_C} shows the evolution of $\Delta_1$ (top panel), $\Delta_2$ (middle panel), and $\Delta \varpi$ (bottom panel),
the quantities defined in Section \ref{approxdelta}, for different values of 
the initial value $\delta.$ Black solid, red dotted, blue dashed, and green dash-dotted curves correspond to $\delta$ = 2.66, 2.71, 2.65, and 2.73 radians, respectively. To verify the validity of our formalism, we calculate the average value of $E/E_\mathrm{crit},$
 which is expected to indicate whether libration or circulation occurs (see discussion below eq.(\ref{cr6})),  over the time interval shown in Fig.~\ref{fig4_C}. 
 The values we obtained, together with the initial value, $\delta$, are given in the legend. 
The time is normalised by the timescale $t_\mathrm{\parallel}$. 

One can see that $\Delta_1$ exhibits significant variations
on this timescale, while $\Delta_2$ exhibits variations on a much smaller timescale with a much smaller amplitude, 
which is in full agreement with our analytic results discussed in Section \ref{critical_curves}.  One can also see that
the cases corresponding to $E/E_\mathrm{crit} < 1$  lead to librations of the apsidal angle, $\varpi$, while the case with
 $E/E_\mathrm{crit} > 1$  is  circulating. In agreement with the theory, the magnitude of the variations of $\delta_k$ increases when moving from circulating to librating solutions.

\subsubsection{Case B}\label{caseB}

Case B represents a close-in system with a near-polar orbit. 
In this case, the stellar parameters correspond to the models of DI Her described above. 
The initial rotational periods of the components are fixed at $1.1$ days, 
in accordance with the rotation period of DI Her A. 
The orbital separation is $0.08$ AU, and the initial eccentricity is 0.5. 
The initial value of the apsidal angle $\varpi$ is 5.7 radians  ($326^{\circ}$), which is also in agreement with the latest estimates for DI Her, see \cite{Liang}. For this case, the value of the parameter $\xi$ is close to $0.72$.

This choice of initial parameters allows us to look at the evolution of the system near the critical curve in the vicinity of $\delta \approx 1.7$. 
The results illustrating the orbital evolution for Case B are shown in the right panel of Fig.~\ref{fig4_C}. 
One can see that there are two cases  undergoing libration. 
Due to the decrease in the orbital separation and increase in stellar rotation rates compared to case A, the librations are more rapid, with the timescale $t_\mathrm{\parallel}$ being reduced significantly. 
The latter is $6.5 \times 10^5$ yrs and $1.8 \times 10^3$ yrs in cases A and B, respectively. Furthermore, for case B,
the timescale $t_{\perp}$ associated with the perpendicular torque is longer than the corresponding timescale for case A. These timescales are
$1.2 \times 10^4$ yrs and $3.8 \times 10^4$ yrs for cases A and B, respectively. 
Thus, contrary to the situation for case A, 
the timescale $t_\mathrm{\parallel}$  no longer governs the 'slow' evolution since $t_\mathrm{\parallel} < t_{\perp}$. 
This results in qualitatively new features in the behaviour of the numerical solutions, which are not described by our simple analytic treatment.
 
Firstly, one can see the absence of small-scale high-frequency variations in $\varpi$ driven by the oscillations of $T_{\perp}$,
 which are clearly seen in the bottom left panel. At the same time, the amplitude of the oscillations in $\varpi$ is substantially larger. Another qualitative difference occurs in the evolution of $\delta_k$. When the state of the system is such that it is displaced slightly from the critical curve, 
both $\Delta_1$ and $\Delta_2$  show variations with a comparable amplitude.
It is interesting to point out, that although our theory developed in \ref{critical_curves} is not applicable in this case,
 there are librating solutions for two of the values of $\delta$ considered, albeit with larger amplitudes than found for case A. 
 The amplitude of variations of $\Delta_{1,2}$ is 
also of the order of $\Delta^\mathrm{max}$.
\begin{figure*}
\begin{multicols}{2}
\includegraphics[width=\linewidth]{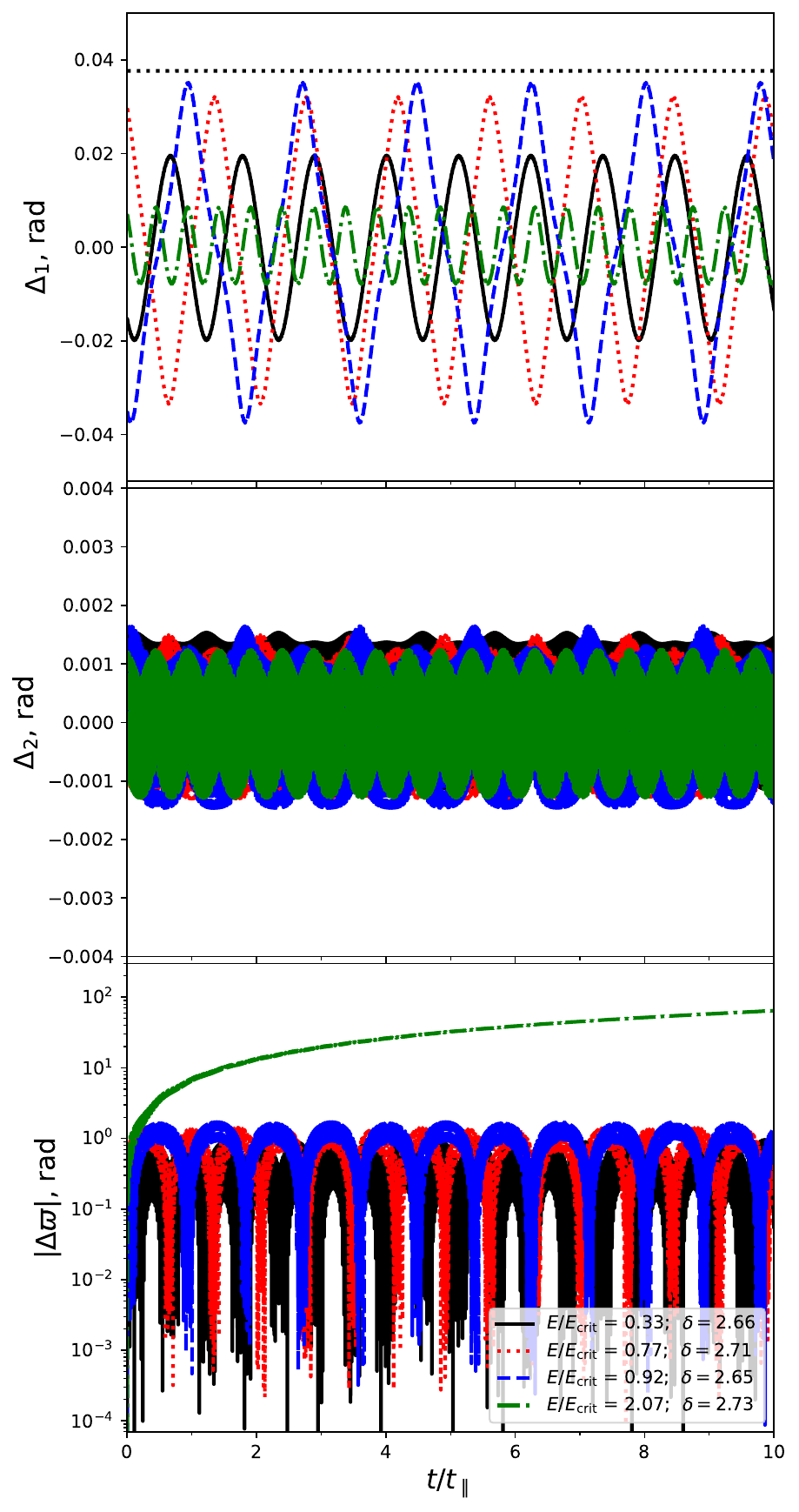}\par 
\includegraphics[width=\linewidth]{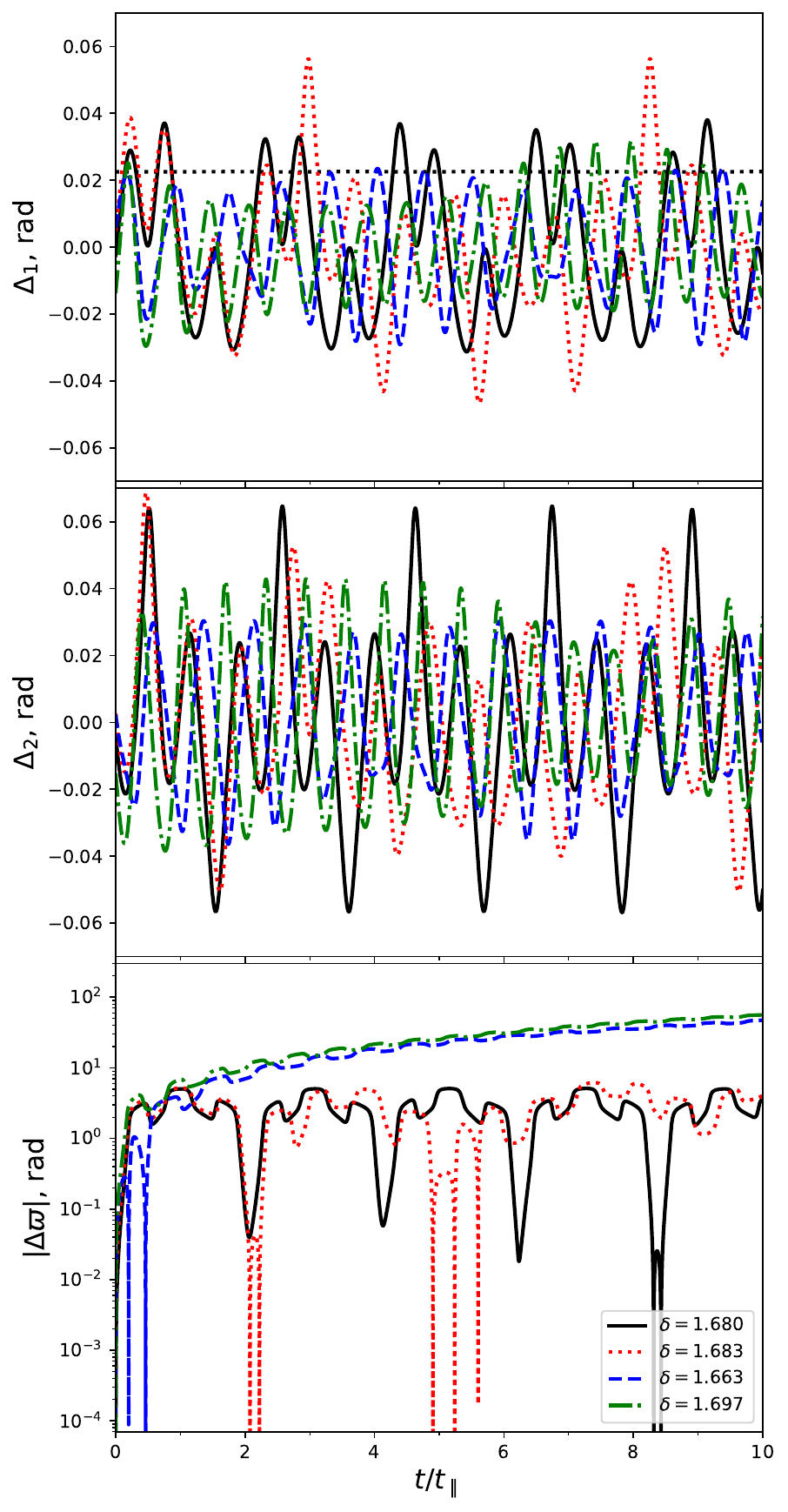}\par
\end{multicols}
\caption{The orbital evolution of four systems in the vicinity of a critical curve. The left panels correspond to case A
and the right panels to case B. The values in the legends are the average values of $E/E_\mathrm{crit}$  over the time interval 
illustrated and the initial values of $\delta$. Top panel and middle panel: the evolution of $\Delta_1$ and $\Delta_2$ is shown.
Grey dotted lines indicate $\Delta^\mathrm{max}$ 
calculated from eq.~(\ref{cr8}). The bottom panels: the evolution of $\Delta \varpi$ is shown. Bottom left panel: Case A,  bottom right panel: Case B (see the text for more detail).}

\label{fig4_C}
\end{figure*}}

{ \section{Summary and Conclusions}\label{conclusions}
\subsection{Basic governing equations}
In this Paper, we have derived a  set of equations describing the general  non-dissipative evolution of
the orbit of a binary system together with the orientations of the spin axes of the two components resulting from their mutual tidal interaction.
Unlike in IP,  IP1, and IP2,  neither of these is assumed to be compact.
This leads to the complicating feature that individual spin angular momenta are no longer 
 in the same plane as the orbital and total angular momentum vectors.
The orbit is allowed to have arbitrary eccentricity.

The tidal torques acting in the system originate from the rotational distortion of the components
{  \citep[see e.g.][]{BOC}} as well as from the recently discovered contribution determined by the effect of Coriolis and inertial forces on the tidal responses of each component (see IP, IP1, and IP2).  
    The torque arising from the rotational distortion is always perpendicular to the plane containing the unit vector in the direction of the orbital
angular momentum and the unit vector in the direction of 
the stellar spin. While the torque arising from inertial effects and Coriolis forces has in addition to a component in this direction a component parallel to that plane.
The latter can lead to qualitatively new dynamics and we describe it as the parallel torque.
The total torque is then written as the sum of the parallel torque and the perpendicular torque.
Explicit expressions for these components are given in Appendix \ref{app1}.
The parallel torque depends on the position of the apsidal angle. Thus, to fully describe the system, we consider all effective contributions to the apsidal precession rate that
operate for an isolated binary \citep[see][and Appendix \ref{app2}]{BOC}. 

The complete system of equations governing the evolution  are given by \\
 eqns ~(\ref{eq14})--(\ref{eq18}) and eq. (\ref{prec}) (see Section \ref{finaleq}). 
 These require specification of the semi-major axis and the magnitude of the angular velocities of each component, which are conserved, as well as the initial values for the five angles specifying the orientations of the spins, the orientation of the apsidal angle, and the orbital eccentricity.
\subsection{Evaluation of the tidal torques for particular stellar models}

Following  IP and IP1, in Appendix \ref{appendixA},
we show how to obtain all quantities characterising the parallel torque in terms of the eigenfunctions and eigenfrequencies
of the normal modes of stellar pulsation and calculate them explicitly for a set of stellar models that are appropriate for the components of the DI Her binary system.  We obtain these models using the \texttt{MESA} stellar evolution code \citep[see][]{MESA1,MESA2,MESA3,MESA4,MESA5}.

\subsection{The case of small orbital inclination}

For most binary systems, it is expected that the ratio of the magnitude of the spin angular momentum to the magnitude of the orbital angular momentum for both components is small. Accordingly, we
 adapt our dynamical system to this case (see Section \ref{aqual})  
and find approximate analytic solutions.
\subsection{Approximate analytic treatment}
 In Sections \ref{approxdelta}--\ref{critical_curves}, it is shown that, when the spin inclination angles $\delta_k$ are not very close to $\pi/2$, the angle determining the relative orientation
of the spin vectors with respect to a fixed orbital frame, $\nu_1-\nu_2,$ exhibits steady rotation under the action of the perpendicular torques. Under the same approximation, the
 angles specifying the inclination to the total angular momentum
  undergo oscillatory variations due to both parallel and perpendicular torques that are characterised by two different timescales. The amplitudes 
of these variations are typically small.

 The parallel torques produce a larger amplitude near the so-called critical curves, where the rate of evolution
of the apsidal angle (defined with respect to the line of nodes determining the intersection of the planes perpendicular to the orbital and total angular momentum) passes through zero.
 Under the assumption that the timescale associated with the perpendicular torques
   is much smaller than the one associated with the parallel torques, we perform a time averaging to obtain equations governing the secular rate 
   evolution of the apsidal angle and spin inclinations, obtaining critical curves from the condition that the former is zero.
 
 Close to such curves,  as is the case when one of the components is a point-like source of gravity (see IP1),
  there are regions in the phase space of the system where the apsidal angle can demonstrate either librating or circulating behaviour.
In particular, the motion of inclination angles can be approximately described in terms of a standard simple pendulum subject to high-frequency forcing produced by the action of the perpendicular torques.
This forcing can lead to chaotic behaviour of the system in the neighbourhood of a separatrix.

\subsection{ Numerical calculations }

We calculate  the locations of  critical curves for systems 
 in the  $(\tilde a, \sigma)$  plane for a range of values of  $\delta_1^0= \delta^2_0\equiv \delta$,
  and in the  $(\tilde a, \delta)$ planes for a range of values of  $\Omega_1/n_0=\Omega_2/n_0 \equiv \sigma.$
We also calculate the rate of precession of the longitude of periapsis on the critical curves. 

We considered the orbital evolution for two representative cases. 
 Parameters corresponding to case A were chosen in such a way that
the assumptions underlying our theoretical approach described in Section \ref{aqual}  are valid,
and we expect $\varpi$ to undergo either circulation or libration in the neighbourhood of a critical curve. 
These are modulated by high-frequency oscillations on a shorter timescale.
  The results of numerical calculations appear to be in good agreement with our theoretical treatment. In particular, we managed to observe the transition between librating and circulating solutions, when the ratio $E/E_\mathrm{crit}$ crosses unity. Moreover, we found that the obliquity variation of the component with a higher absolute value of the projection of the spin vector onto the plane perpendicular to ${\bf J}$, $\Delta_1$, follows the restriction laid by (\ref{cr8}). The second component exhibits high-frequency changes in $\Delta_2$ which appear negligible compared to the first component, also in agreement with our theory.  }
  
{ It is of interest to rewrite $\Delta^\mathrm{max},$ given by (\ref{cr8}),  under the  assumptions 
of systems with comparable masses, inclinations, and structure of the components. 
We obtain
\begin{equation}
\Delta^{max} \sim \sqrt{\left((1-\xi^2) {18(1+2\beta_{*})e^2(1+{e^{2}\over 6})\over {\tilde \omega_\mathrm{eq}}^2{\tilde r_p}^{3}(1+e)^3}\right)},
\label{conc1}
\end{equation}
where $\tilde r_p=(1-e)a/R$ is dimensionless periastron distance. 
Equation (\ref{conc1}) indicates that it is  difficult to obtain large variations of $\delta$, with $\Delta^{max} >\sim 1$,
for systems with sufficiently large values of $\tilde \omega_\mathrm{eq}$, as in our models of DI Her, for which  $\tilde \omega_\mathrm{eq}\sim 7$. 
Using $\beta_{*}=0.7$, $\sqrt{(1-\xi^2)} \sim 1$ 
and $\tilde \omega_\mathrm{eq}\sim 7$, we obtain $\Delta^{\max} \sim 0.9e(1+{e^{2}/6})^{1/2} (1+e)^{-3/2}{\tilde r_p}^{-3/2}$. Note, however, that this estimate
does not apply to nearly polar orbits as considered in case B and stars with unequal parameters.

  Case B illustrates a system with parameters close to those suggested for DI Her,
   apart from the angle $\delta$, which is initiated slightly above $\pi/2$ to have a retrograde and nearly polar orbit
    with its parameters close to ones corresponding to a system on a critical curve.   
     We recall our discussion at the beginning of Section \ref{critDIHer}, which points out that the fact that the secondary component is close to retrograde rotation suggests the possibility of libration about a centre located on a critical curve as found in Case B. Although the assumptions of our simple analytical approach are not valid for these simulations, it still gives some qualitative agreement with the
numerical results. In particular, close to a critical curve, libration of the apsidal angle is observed, while further away circulation occurs.
 In addition, the maximum amplitude 
of variations of $\delta_k$ is close to the maximal theoretical values $\Delta^\mathrm{max}$ given by eq. (\ref{cr8}).  \\

Results reported in this Paper may be used to formulate,  carry out, and interpret
 accurate calculations of the dynamics of misaligned eccentric close binary systems on timescales, which are typically smaller than required for dissipative tidal evolution. A more thorough application to particular systems clearly deserves a separate study.   }

\section*{Acknowledgments}
The work was supported by the Foundation for the Advancement of Theoretical Physics and Mathematics BASIS. We are grateful
to the referee, Michael Efroimsky, for his valuable comments and suggestions.

\section*{Data Availability}
The data underlying this article will be shared on reasonable request to the corresponding author.






\appendix

\begin{appendix}
\section{ Parameters associated with evolved stellar models that are required to quantify the dynamics}\label{appendixA}

\subsection{Evolved stellar models for the components of the DI Her system that are consistent with observed masses, radii for two assumed  metallicities}
\label{Herc}

In order to study DI Her, we use stellar evolutionary models obtained with the  \texttt{MESA} code \citep[see][]{MESA1,MESA2,MESA3,MESA4,MESA5} applying the prescriptions from the MIST project (\citealt{Dotter,Choi}). For each component, we consider three possible mass values corresponding to the mean value as well as the upper and lower limits taken from \cite{Liang}. Noting the chemical abundances provided by \cite{Amard}, we consider two metallicities representative of the range characteristic of the Galactic thin disk population, namely [Fe/H] = -0.3 and +0.3.

 Fig.~\ref{fig0_F} illustrates the evolution of the radius of each component. {The  horizontal green strip highlights the ranges 
 determined by \cite{Liang}, being  [2.71 $R_{\odot},$  2.90 $R_{\odot}$] and [2.49 $R_{\odot},$  2.67 $R_{\odot}]$ }for the primary and secondary component, respectively. 
 One can see the major importance of metallicity as it defines how often and at which evolutionary stage the radius of each component lies within the observed range. Thus, metal-poor stars, depicted with black curves, cross the green area twice during their lifetime,  once shortly before reaching the zero-age main sequence (ZAMS), and again during the
  late main sequence (MS) stage. In contrast, metal-rich stars, depicted with red curves, are characterized by larger radii for a given mass and age and occupy the observed range only for the first 10 Myr of their MS stage.

\begin{table*}
	\centering
	\caption{Parameters of stellar models:\label{tab1}
{The first column contains the model identifier, the second the mass, the third the metallicity, the fourth the age,  the fifth the radius, the sixth  the moment of inertia
	in units of $MR^2$, the seventh, $Q_\mathrm{eq},$  the eighth, $\tilde \omega_\mathrm{eq}$,  and the ninth, $\beta^*$.}}
	
	\begin{tabular}{|c||c|c|c|c|c|c|c|c|} 
		\hline
	   	Model  & $M [M_{\odot}]$ & [Fe/H] & $t$ [yr] & $R [R_{\odot}]$ & $I [MR^2]$ & $Q_\mathrm{eq}$& $\tilde \omega_\mathrm{eq}$ &  $\beta_*$ \\
		\hline
		\texttt{{Primary}}\\
		\hline
		   \texttt{M51PY} & 5.10 & -0.3 &  $5.64 \times 10^5$ & 2.79 & $8.60 \times 10^{-2}$ & 0.358 & 6.75 & 0.706\\
        \texttt{M51PO} & 5.10 & +0.3 & $3.19 \times 10^7$ & 2.77 & $8.62 \times 10^{-2}$ & 0.360 & 6.70 & 0.711 \\
        \texttt{M51RY} & 5.10 & -0.3 & $1.03 \times 10^6$ & 2.79 & $8.38 \times 10^{-2}$ & 0.349 & 6.83 & 0.702 \\
        \texttt{M51RO} & 5.10 & +0.3 & $4.88 \times 10^6$ & 2.79 & $8.59 \times 10^{-2}$ & 0.356 & 6.80 & 0.703\\
\hline
\texttt{{Secondary}}\\
\hline
        \texttt{M44PY} & 4.40 & -0.3 & $8.25 \times 10^5$ & 2.59& $8.29 \times 10^{-2}$ & 0.348 & 6.82& 0.705\\
        \texttt{M44PO} & 4.40 & +0.3 & $4.48 \times 10^7$ & 2.54& $8.29 \times 10^{-2}$ & 0.349 & 6.78 & 0.710\\
        \texttt{M44RY} & 4.40 & -0.3 & $1.48 \times 10^6$ & 2.66& $7.66 \times 10^{-2}$ & 0.325 & 6.90 & 0.700\\
        \texttt{M44RO} & 4.40 & +0.3 & $9.08 \times 10^6$ & 2.59& $8.20 \times 10^{-2}$ & 0.343 & 6.83 & 0.702\\
     		\hline
	\end{tabular}
	\label{table1}
\end{table*}

{{For each mass-metallicity combination, we select two models taken at the moments when a star enters the radius range for the first time (referred to later as a young model) 
 and when a star is about to leave this range for the last time (referred to later as an old model).}

{The main properties of {particular}  models with $M = 5.5 M_{\odot}$ and $M = 4.4 M_{\odot}$ are reported in Table~\ref{table1}.
 In Table \ref{table1},  we purposely focus on results for masses of the components centred within the estimated limits specified in \cite{Liang}. 
 Since, as will be shown later, varying the masses within these limits does not impact our findings.}
 Each model is labeled with an identifier of the form, { \texttt{Ma$\bar{\rm{a}}$bc}, where \texttt{a$\bar{\rm a}$} refers to the stellar mass in solar units, \text{b}  indicates the  chemical abundance 
 (with \texttt{P} indicating the metal-poor and \texttt{R} indicating the metal-rich model), and \texttt{c} indicates the model's age (with  \texttt{Y} and \texttt{O} denoting  a young and old model, respectively).}

These models are illustrated in  Fig.~\ref{fig1_F}. In the upper panels, we show the density normalised by the mean density. It is clear that varying the metallicity, mass, and age within reasonable limits leads to negligible variations in density profiles. In the middle and lower panels, we show the square of the Brunt–Vaisala frequency normalised by the square of the critical frequency $\Omega_* = \sqrt{{GM/R^3}}$ for the young and old models, respectively.} Our models demonstrate pronounced differences in Brunt–Vaisala frequency, especially near the core/envelope boundary, where $N^2$ changes sign, with metal-rich stars, depicted in red, possessing larger cores in general. One can also see that the young models tend to have smaller cores.

\subsection{ Determination of the  overlap integral,   \texorpdfstring{$Q_{\mathrm{eq}}$}{Lg},  the characteristic frequency,  \texorpdfstring{$\omega_{\mathrm{eq}}$}{Lg},  
and the rotational frequency shift parameter, \texorpdfstring{$\beta_{*}$}{Lg}, 
for the equilibrium tide  in terms of a normal mode expansion for a particular stellar model}
\label{sec:method}
Here, we determine expressions for $Q_\mathrm{eq},$ $\omega_\mathrm{eq}$ and $\beta_*$  in the form of eigenfunction expansions over the normal modes of adiabatic oscillation.
We begin by expressing the displacement vector {associated with the equilibrium tide,}  ${\bm{\xi}}_{\mathrm{eq},n,k}$, defined as the solution of eq. (36) of IP
as a sum over the normal modes, ${\bm {\xi}_i}$, with eigenfrequency,  $\omega_i$,  that are associated with the problem of free adiabatic stellar pulsation. Thus we write 
\begin{equation}
{\bm {\xi}}_{\mathrm{eq},n,k}=\sum_{i \ge 1} A_{i}{\bm{\xi}_{i}}.
\label{e1}
\end{equation}
 Using the orthogonality of the  ${\bm {\xi}}_i$  expressed through the condition that \noindent   $\int_V \rho { \bm {\xi}_j}^*\cdot {\bm{ \xi_i}}dV= N_i\delta_{ij},$ \footnote{Note that  as the eigenvalue for the adiabatic pulsation problem depends 
only on $\omega_i^2$ for a spherical star, we may limit consideration
to modes with positive values of $\omega_i$ only.}
(this integral and others below are taken over the volume of the star), we find
\begin{equation}
A_{i} = - A_{n,k} \frac {Q_i}{\omega_i^2N_i}.
\label{e2}
\end{equation}
{Here, the constant} ${\cal A}_{n,k}$, as defined in eq. (30) of IP, is related to the
forcing tidal potential, $U$, through $U = A_{n,k}r^2Y_{2,n}(\theta,\phi)$; $Y_{2,n}$  denotes the spherical harmonic with $l=2$ and azimuthal mode number, $n$,   
\begin{equation}
Q_i=2\int \rho r^3dr(\xi^r_i+3\xi^S_i), \quad {\rm and}\quad
N_i= \int \rho r^2dr(({\xi^r_i})^2+6({\xi^S_i})^2).
\label{e3}
\end{equation}
The radial component of the displacement vector, ${\bm{\xi}}_i$,  is $\xi_i^r Y_{2,m}$, and $r\xi_i^{S}\nabla_{\perp}Y_{2,m}$ gives the angular components (see eq. (27) of IP). All other symbols have the usual meaning.

Using eqns (\ref{e1}) and (\ref{e2}), it is straightforward to 
calculate the norm $N_0$, the frequency $\omega_\mathrm{eq}$, and the normalised overlap integral
$Q_\mathrm{eq}$, defined in eqns (46), (47), and (59) of IP, respectively. We obtain {
\begin{equation}
N_0={\cal A}_{n,k}^{2}\sum_{i \ge 1} \frac {Q^2_i}{\omega_i^4N_i},
\label{e4}
\end{equation}
\begin{equation}
\omega_\mathrm{eq}^2 =\left(\sum_{i \ge 1} \frac {Q^2_i} {\omega_i^2N_i}\right)\left( \sum_{i \ge 1} \frac {Q^2_i}{\omega_i^4N_i}\right)^{-1}
\label{e5}
\end{equation} 
and
\begin{equation}
\frac{Q_\mathrm{eq}^2}{N_0}=\left(\sum_{i \ge 1} \frac {Q^2_i} {\omega_i^2N_i}\right)^2\left(\sum_{i \ge 1} \frac {Q^2_i} {\omega_i^4N_i}\right)^{-1}.
\label{e6}
\end{equation}

It is easy to see that { $\omega^2_\mathrm{eq}$}  and $Q_\mathrm{eq}$ have the correct dimensions. However, from now on we find it convenient to normalise $\omega_\mathrm{eq}$ by the critical frequency $\Omega_{*}$. Thus, we move from $\omega_\mathrm{eq}$ to $\tilde \omega_\mathrm{eq} = \omega_\mathrm{eq}/\Omega_{*}$.

Similarly, we calculate the coefficient $\beta_*$ determining the frequency shift due to rotation and defined in eq. (55) of IP which takes the form
\begin{align}
n(\beta_*-1)=
-\frac{ \rm{i}\int \rho {\bm {\xi}}_{eq,n,k}^*\cdot(\hat{\bf k}\times {\bm{\xi}}_{eq,n,k})dV}
 {\int \rho |{\bm {\xi}}_{eq,n,k}|^2dV},\label{IP51}
\end{align} 
with the notation as in IP.
We substitute eqns (\ref{e1}) and (\ref{e2}) to r.h.s. of {(\ref{IP51}) to obtain
\begin{equation}
n(\beta_*-1)=-i\left( \sum_{k,j\ge 1}\frac{\hat Q_k \hat Q_j {S_{kj}}}{ \omega_k^2 \omega_j^2\sqrt N_k \sqrt N_j}\right)   \left( { \sum_{k \ge 1}{\hat Q^2_k\over \omega_k^4}}\right)^{-1},\label{e34}
\end{equation}  }
where we {make use of} eq. (\ref{e4}),
\begin{equation}
\hat Q_k=Q_k/\sqrt N_k, 
\label{e8}
\end{equation}  
and
\begin{equation}
S_{kj}=\int \rho dV {\bm {\xi_{k}}}^{*}\cdot (\hat {\bf k} \times {\bm \xi}_{j}), 
\label{e9}
\end{equation}  
{recalling that  $\hat {\bf k}$ } is the unit vector in the direction of the rotation axis.

Expressing  the eigenfunctions  ${\bf{\xi}}_i$ in terms of $\xi^r_i,$ and $\xi_i^S$ in eq. (\ref{e9}),  a straightforward calculation gives
\begin{equation}
S_{kj}=-in\sigma_{k,j}, \quad {\rm{with}} \quad \sigma_{k,j}=\int r^2dr \rho (\xi^{r}_{k}\xi^{S}_{j}+ 
\xi^{r}_{j}\xi^{S}_{k}+\xi^{S}_{k}\xi^{S}_{j}). 
\label{e10}
\end{equation}  
Substituting eq. (\ref{e10}) in eq. (\ref{e34}), we finally  obtain 
\begin{equation}
\beta_{*}=1-\left(\sum_{k,j}{\hat Q_k \hat Q_j \over \omega_k^2\omega_j^2}{\sigma_{k,j}\over \sqrt N_k \sqrt N_j}\right)\left(\sum_{k\ge 1} \frac{\hat Q^2_k} {\omega_k^4}\right)^{-1}.
\label{e11}
\end{equation}  

\subsection{Calculation of the adiabatic normal modes of oscillation and determination of \texorpdfstring{$Q_\mathrm{eq}, \omega_\mathrm{eq}$}{Lg} and \texorpdfstring{$\beta_*$}{Lg}}
\label{calculations}

 We use the \texttt{GYRE} code (\citealt{GYRE}) to find the free stellar  {adiabatic f mode,}  g-modes and p-modes, 
together with their associated eigenfrequencies and eigenfunctions. These eigenfunctions }are used to compute overlap integrals following the procedure described in Appendix~\ref{sec:method}.
 Fig.~\ref{fig2_F} shows $\omega_\mathrm{eq}$, $Q_\mathrm{eq}$, and $\beta_*$ as a function of the highest radial order $N_\mathrm{r}$ used for calculation.
Increasing $N_\mathrm{r}$ by  $1$ {has the effect of adding in the next order p-mode together with the next order  g-mode starting with the $f$ mode when $N_r=0.$}

As $N_r$ increases,   $Q_\mathrm{eq}$, $\omega_\mathrm{eq}$, and $\beta_*$ {effectively converge for $N_r > 10$}. Accordingly, the values of these quantities given in Table~\ref{tab1} are obtained with $N_\mathrm{r} = 10$. {Tabulated} values of $Q_\mathrm{eq}$ and $\omega_\mathrm{eq}$ are centred around 6.8 and 0.35, respectively,
with the variation being around  
3\% in each case. Tabulated values of $\beta_{*}$, also obtained with $N_\mathrm{r} = 10$, exhibit the smallest relative scatter for the models considered, being less than 2\%.

\begin{figure*}
\begin{multicols}{2}
  \includegraphics[width=\linewidth]{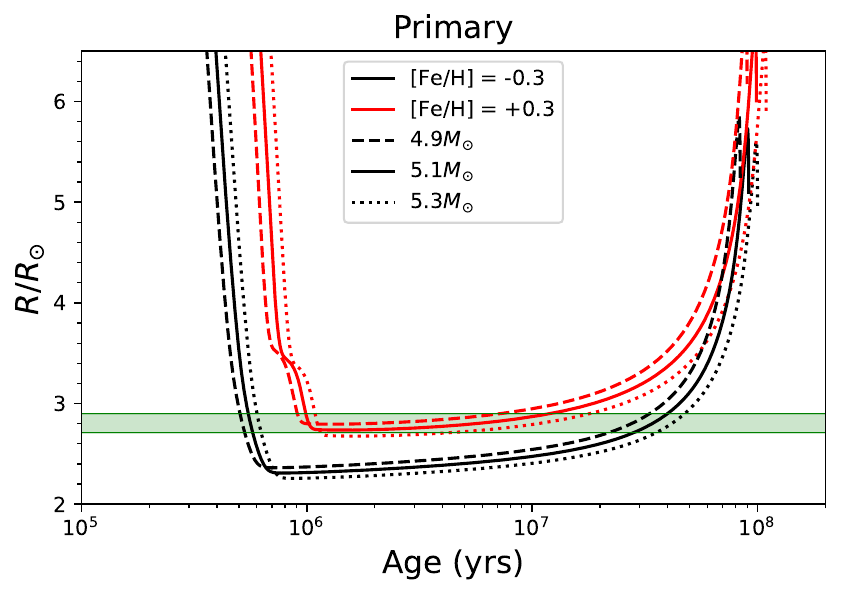}\par 
    \includegraphics[width=\linewidth]{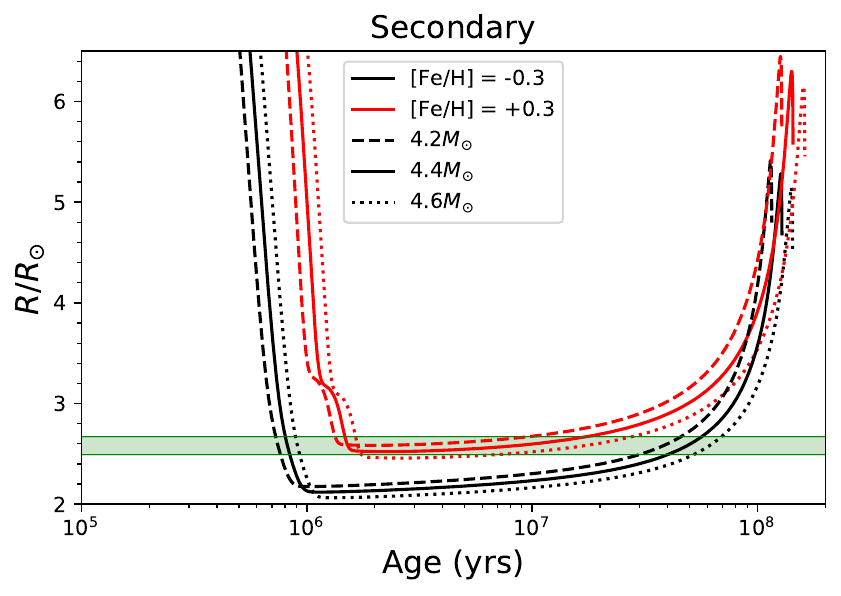}\par
\end{multicols}
    \vspace*{-6mm}
\caption{Evolution of the stellar radii. 
{The left panel shows the evolution of  the  primary component and the right panel the secondary component.}
Black and red curves correspond to low-metallicity and high-metallicity models, respectively.
 Solid lines illustrate the evolution of the model with the mean estimated mass value,  while the dashed (dotted) curves refer to the models with lower (upper) limits for the estimated
 value of the mass.  The area shaded in green indicates the constraints on the radii obtained by~\protect\cite{Liang}. }
\label{fig0_F}
\end{figure*}

\begin{figure*}
\begin{multicols}{2}
    \includegraphics[width=\linewidth]{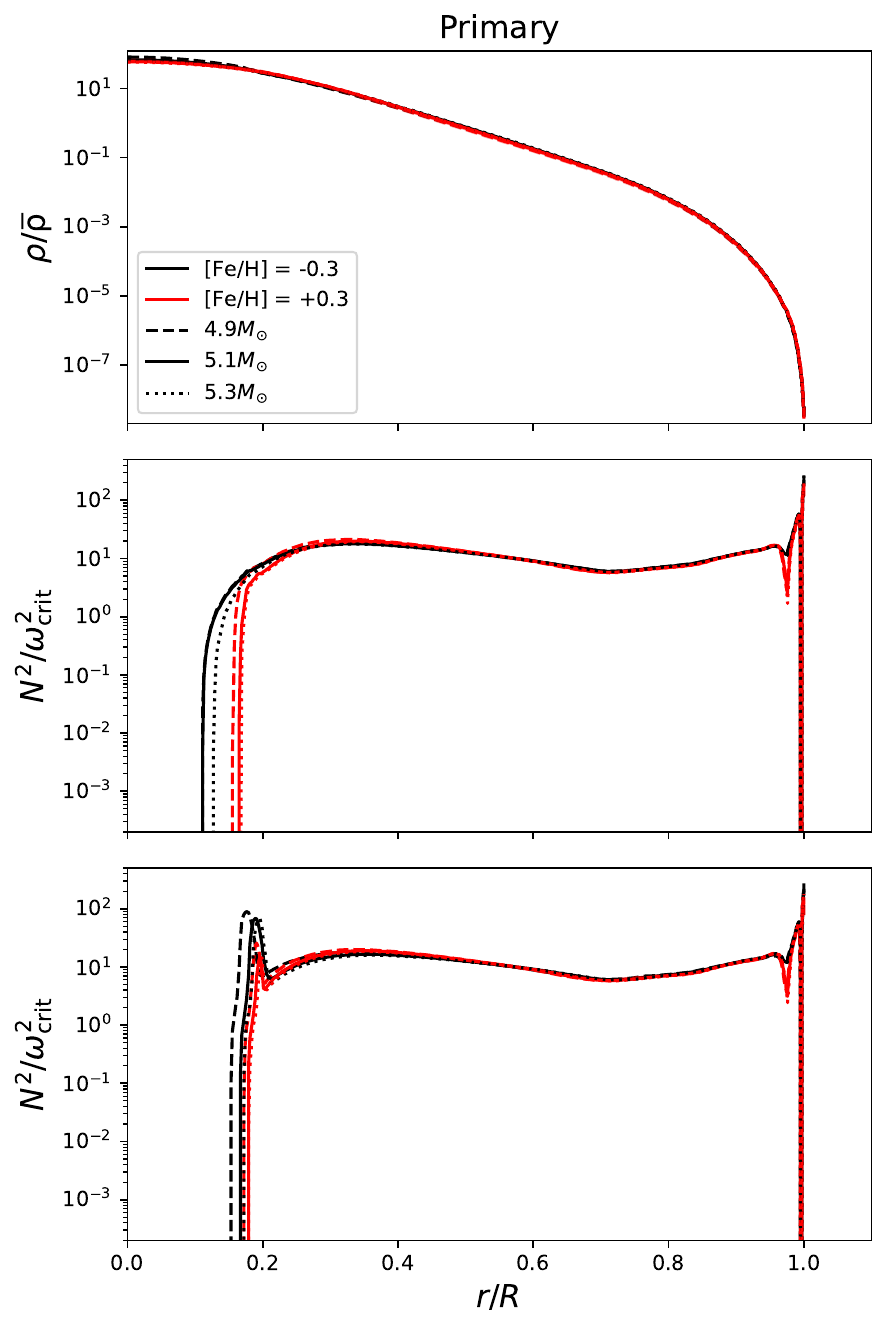}\par 
    \includegraphics[width=\linewidth]{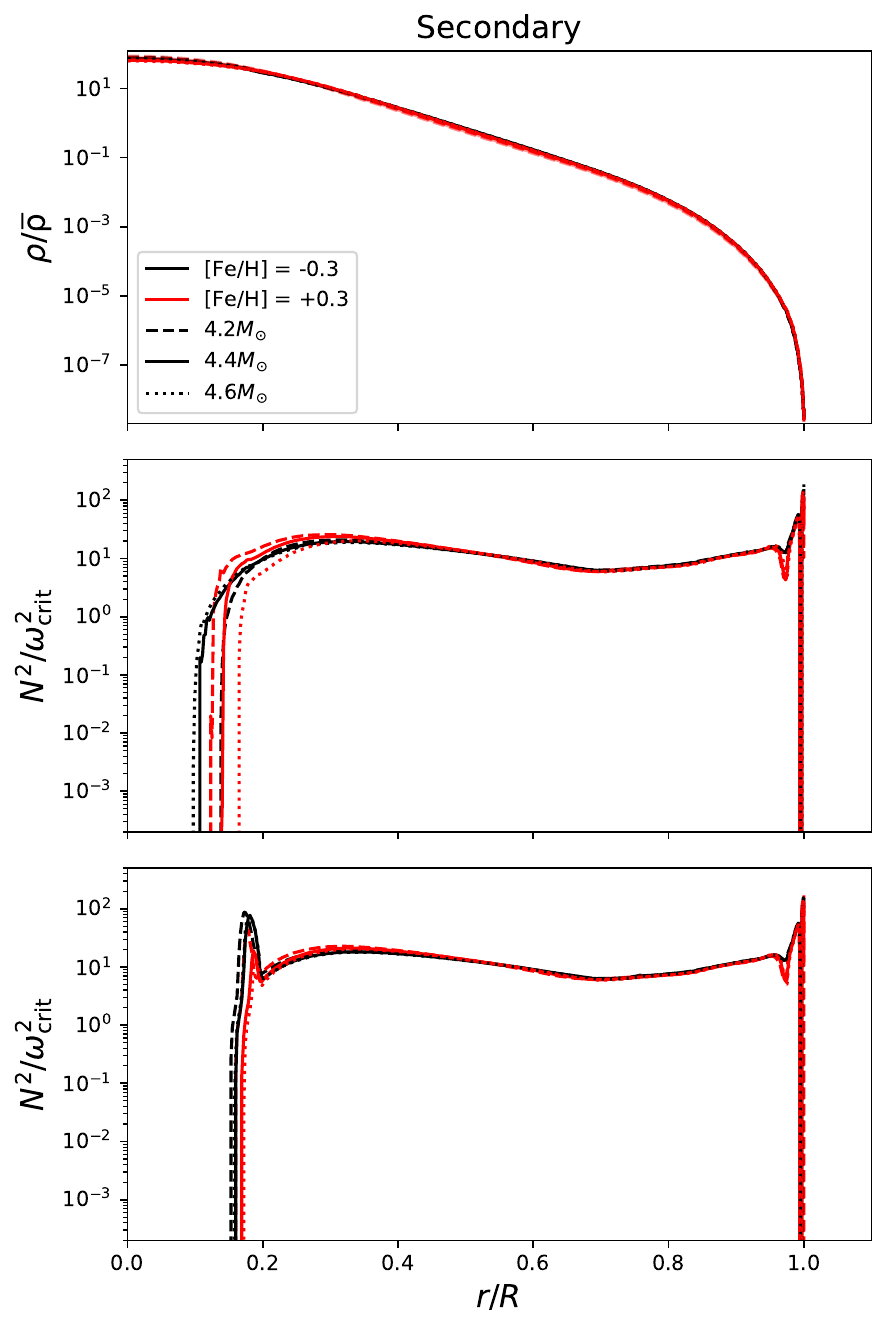}\par 
    \end{multicols}
\caption{Normalised density profiles  (upper panels) and squared Brunt–Vaisala frequency profiles for the lowest (middle panels) and
highest (lower panels) age models of the components of DI Her. Designations are the same as in Fig.~\ref{fig0_F}.}
\label{fig1_F}
\end{figure*}

\begin{figure*}
\begin{multicols}{2}
    \includegraphics[width=\linewidth]{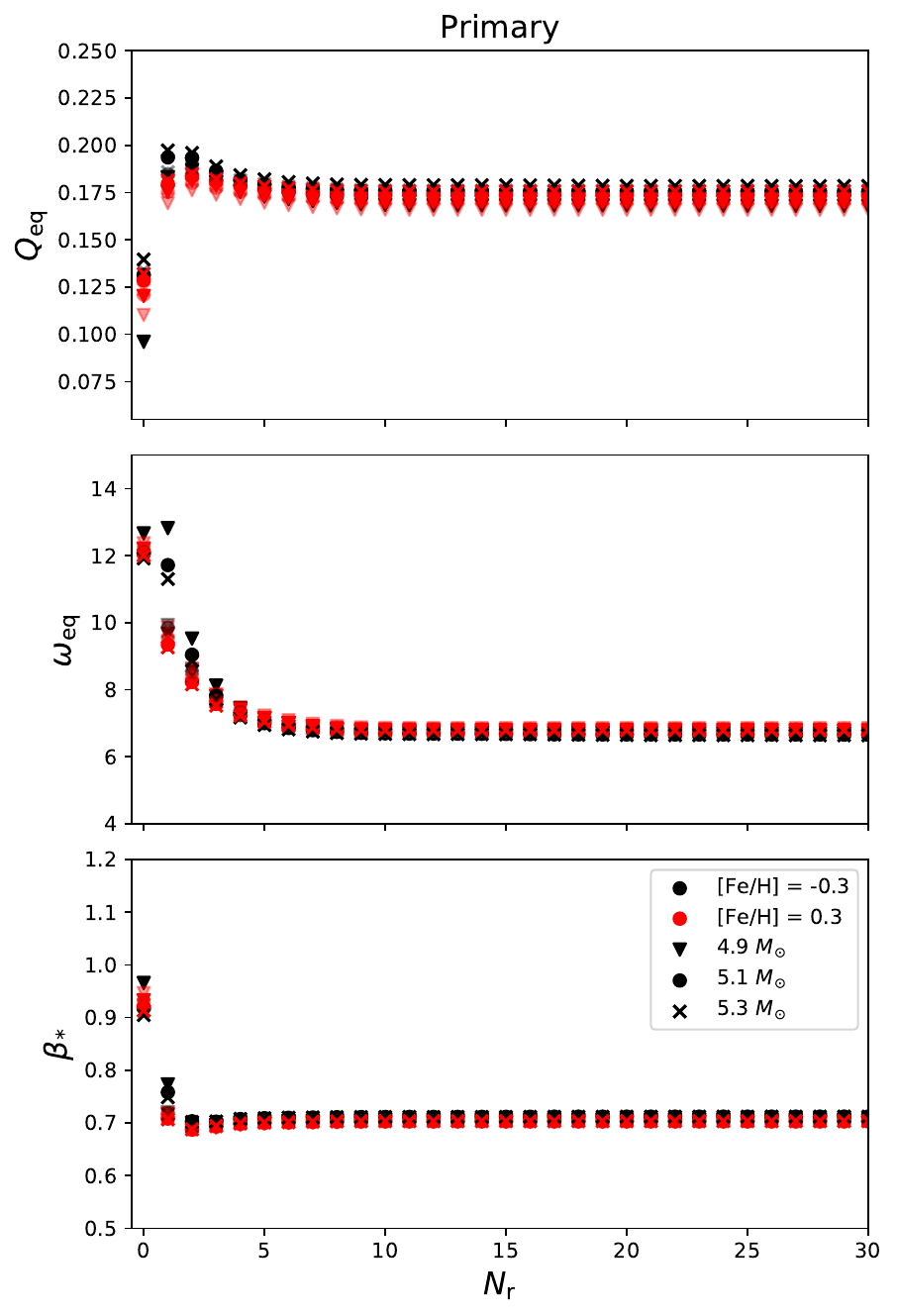}\par 
    \includegraphics[width=\linewidth]{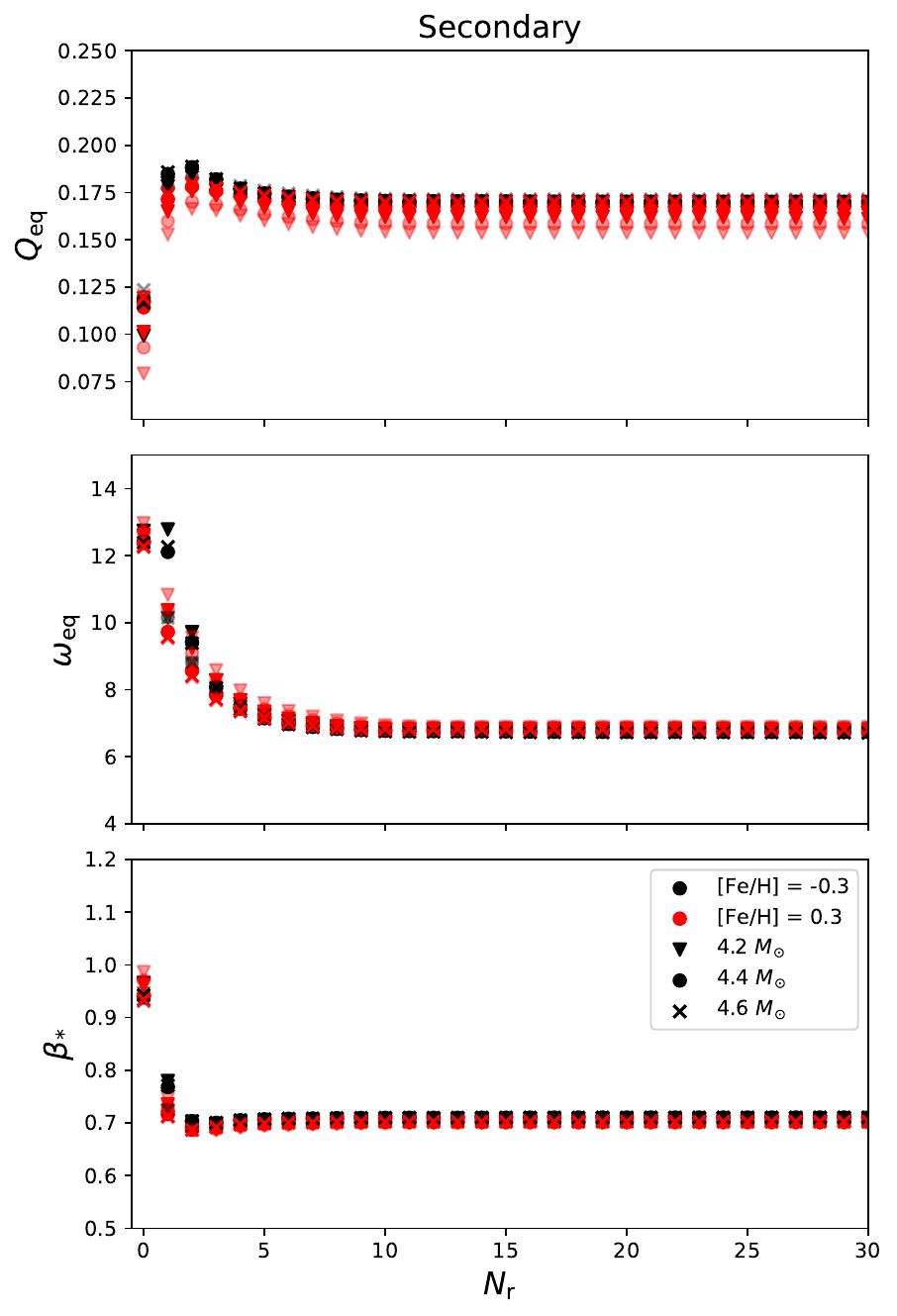}\par 
    \end{multicols}
\caption{The overlap integral
$Q_\mathrm{eq}$ (upper panel), normalised frequency $\omega_\mathrm{eq}$ (middle panel), and coefficient $\beta_*$ (bottom panel) computed for the models shown in Fig.~\ref{fig1_F} as a function of the highest radial order $N_\mathrm{r}$ used for calculation. For a given symbol, fainter colouring represents younger models, while brighter colouring represents older models. }
\label{fig2_F}
\end{figure*}

\section{The explicit expressions for the torque components \texorpdfstring{$T_{\parallel,k}$}{Lg} and \texorpdfstring{$T_{\perp,k}$}{Lg}}
\label{app1}

Noting the discussion of the coordinate systems in Section \ref{subsec:rel}
and eq. (\ref{eq6}), it can be seen that  the torque components $T_{\parallel,k}$ are given by the expressions
(90) and (92) of IP as \footnote{Note the following misprints in table 2 of IP, ${\hat \Omega}$ should be replaced by $\tilde\Omega,$ which should equal  $n_0/\omega_\mathrm{eq}$.
}
\begin{align}
&T_{\parallel,k}= \nonumber\\&-T_{*,k}   \frac{3(2\beta_{*,k}+1)}{2} e^2(1-e^2)^{3/2}\left(1+\frac{e^2}{6}\right)
\left( \frac{\Omega_{r,k}}{\omega_{eq,k}} \right)^2\sin\beta_k\sin{2{\hat \varpi_k}},  \nonumber \\
&\hspace{-3mm}{\rm with}\hspace{2mm}T_{*,k}=\frac{6\pi}{5}\left(\frac{ GM_{j}Q_{eq,k}}{ a^3(1-e^2)^3\sqrt{N_{0,k}}\omega_{eq,k}}\right)^2  = \nonumber\\ &
 \frac{3k_{2,k}M^2_{j}}{ M_\mathrm{T}} \left(\frac {R_{*,k}}{a}\right)^5{ n_{o}^{2} a^2}{(1 - e^2)^{-6}}.\hspace{3mm}
 \label{TX}
\end{align}
Here we remind that $e$ and $a$ are orbital eccentricity and semi-major axis, $n_0$ is 
the mean motion, $M_k$,
$R_{*,k}$, and $\Omega_{r,k}$ are the mass, radius, and rotational frequency of the $kth$ component, respectively, and $j\ne k$.
The total mass of the system is $M_\mathrm{T}= M_1+M_2$ and the last equality is regarded as defining the apsidal motion constant $k_{2,k}$. The quantities
$Q_{eq,k}$, $N_{0,k}$, $\omega_{eq,k}$, and $\beta_{*,k}$
correspond to the quantities $Q_\mathrm{eq}$, $N_{0}$, $\omega_\mathrm{eq}$, and $\beta_*$
that are evaluated as described in Section \ref{sec:method} for the component denoted with the subscript $k.$

An expression for the torque component $T_{\perp}$, convenient for our purposes, can be obtained from finding $T^y$ from eq. (90) of IP and then adding in the torque $T^y_\mathrm{SF}$, given by eq. (D9) of IP. Doing this, we obtain\footnote{Note that this procedure was followed to obtain eq. (95) of IP, in which there is a misprint.
A factor $\lambda^2$, that should have been in the last term of that equation, was omitted.}

\begin{align}
T_{\perp,k}=& -2T_{*,k}\sin\beta_k {\left(\frac{\Omega_{r,k}n_0}{\omega_{eq,k}^2}\right)}\Biggl(\phi_1\beta_{*,k}-\nonumber\\&(1-e^2)^{3/2}\frac{\Omega_{r,k}}{n_0}(\beta_{*,k}+1/2)
(\phi_2+\phi_3\cos2\hat \varpi_k)\cos\beta_k\Biggr)+\nonumber\\
&T_{*,k}\frac{M_\mathrm{T}}{{3}M_{j}}\left(\frac{\Omega_{r,k}}{n_0}\right)^2(1-e^2)^{9/2}\cos\beta_k\sin\beta_k, \label{Tperp}
\end{align}
where as given in by eqns (27)-(29) of IP, \\  $\phi_1 = 1 +15e^2/2 + 45 e^4/8 +5e^6/16,\\
 \phi_2 = 1 + 3e^2 +3e^4/8,$ and $\phi_3= 3e^2/2+e^4/4.$

The angles $\hat \varpi_k $ measure the locations of the apsidal line with respect to the line of intersection of the equatorial plane of a particular component and the orbital plane, see Section \ref{subsec:rel}, \cite{IP}  and \cite{IP1}. 
They are related to the angle $\varpi$ measuring the location of the apsidal line with respect to the line of nodes located along the intersection of the plane perpendicular to the total angular momentum, ${\bf J}$ with the orbital plane, as $\hat \varpi=\varpi +\gamma_k$, where $\gamma_k$
are defined in eq.(\ref{gammak}), see Section \ref{subsec:rel}. 

Since $\gamma_k$ are solely determined by the variables, whose dynamical evolution
is already described above, we need only consider  the evolution of $\varpi$, when determining  the evolution of $\hat \varpi_{k}$ due
to the influence of the various effects causing apsidal precession.
This is done in Section \ref{app2}.

\section{Apsidal precession resulting from the rotational flattening of the two stars:
Rotational distortion and non-inertial effects  resulting from precession of the stellar spins}\label{app2}
\cite{BOC}  derived a general expression for the joint precession of the orbital angular momentum and the eccentricity vector, 
{ ${\bf A} =e{\mbox{{\boldmath$\epsilon$}}}$, where ${\mbox{{\boldmath$\epsilon$}}}$ is the unit vector lying along the semi-major axis and pointing towards
the periastron of the orbit}
in the form
\begin{equation}
\dot {\bf L}={\bf \Omega}\times {\bf L}, \quad \dot {\bf A}={\bf \Omega}\times {\bf A},
\label{b1}
\end{equation} 
where ${\bf \Omega}={\bf \Omega}_1 +{\bf \Omega}_2 $  has contributions from both binary components as indicated. For each of these, we have (see IP1)
\begin{equation}
{\bf \Omega}_k ={\Omega}_{s,k}{\bf s}_k+{ \Omega}_{l,k}{\bf l}, \quad \Omega_{s,k}=\Omega_{Q,k}\cos \beta_k, \quad \Omega_{l,k}=\Omega_{Q,k}{1-5\cos^2\beta_k\over 2}{\bf ,}
\label{b2}
\end{equation}
where
\begin{equation}
\Omega_{Q,k}=-k_{2,k}\left(1+\frac{M_{j}}{M_k}\right){\sigma^2_k\over (1-e^2)^2}{\tilde a}^{-5}n_{0},
\label{b3} 
\end{equation}
where $\sigma_k = \Omega_{r,k}/n_0.$   
Noting that eq. (\ref{b1}) does not change the magnitude of ${\bf L}$ or ${\bf A}$, we derive
\footnote{We remark that, if the magnitude of these vectors are allowed to change on account of other causes, it can be checked that the results in this Appendix are not changed. This is because the changes in the vectors calculated correspond to
rotations independent of their magnitudes.}
\begin{equation}
\dot {\bf l}={\bf \Omega}\times {\bf l}, \quad \dot {\mbox{{\boldmath$\epsilon$}}}={\bf \Omega}\times {\mbox{{\boldmath$\epsilon$}}}. 
\label{a1}
\end{equation}

In order to obtain an explicit expression for the apsidal precession rate, it is convenient to use the orthonormal 'orbital' right-handed frame connected to the vector ${\bf j}_{\perp}$ given by eq. (\ref{oy''}) and specified by the unit vectors  $({\bf n}_1, {\bf n}_2, {\bf n}_3)$
\begin{equation}
{\bf n}_{1}\equiv {\bf j}_{\perp}={{\bf j}\times {\bf l}\over \sin i}, \quad {\bf n}_{2}={({\bf l}\times {\bf j})\times {\bf l}\over \sin i}={{\bf j}-\cos i {\bf l}\over
\sin i} , 
\quad {\bf n}_{3}={\bf l}.    
\label{en5}
\end{equation}  
 A similar frame was employed for the calculation of the precession rate for a single active component in \cite{IP1} (see their eq. (8)).

We decompose  ${\mbox{{\boldmath$\epsilon$}}}$ as a linear combination of ${\bf n}_{1}$ 
and ${\bf n}_{2}$, which both lie in the orbital plane:
\begin{equation}
{\mbox{{\boldmath$\epsilon$}}}=\cos \varpi {\bf n}_{1}+\sin \varpi {\bf n}_{2}.
\label{gam5}
\end{equation} 
It is clear that by definition $\varpi$ does not depend on the component considered.

 Now we differentiate (\ref{gam5}) with respect to time to obtain
\begin{equation}
\dot {{\mbox{{\boldmath$\epsilon$}}}}=(\cos \varpi {\bf n}_{2}-\sin \varpi {\bf n}_{1})\dot \varpi 
+\cos \varpi \dot {\bf n}_{1}+\sin \varpi \dot {\bf n}_{2}.
\label{a10}
\end{equation}

In order to obtain an expression for the apsidal precession rate, we take the scalar product of eq. (\ref{a10}) with ${\bf n}_{1}$. Then we substitute (\ref{a10}) in (\ref{a1}) 
and use the known properties of the scalar triple product and the orthogonality of (\ref{en5}) to get
\begin{equation}
{\bf n}_{1}\cdot ({\bf \Omega}\times {\bf n}_{2})=-\dot \varpi + {\bf n}_{1}\cdot \dot {\bf n}_{2}.
\label{p1}
\end{equation}  
The term on l.h.s. can be transformed to $-({\bf \Omega}\cdot {\bf l})$, taking into account that 
${\bf l}={\bf n}_{1}\times {\bf n}_{2}$. Accordingly, we have
\begin{equation}
\dot \varpi = ({\bf \Omega}\cdot {\bf l})  + {\bf n}_{1}\cdot \dot {\bf n}_{2}.
\label{p2}
\end{equation}  
Now we use the explicit expression  
(\ref{en5}) 
for $ {\bf n}_{2}$ to express its time derivative through $\dot {\bf l}$, use (\ref{a1})
and the known properties of vector triple product to have  ${\bf n}_{1}\cdot \dot {\bf n}_{2}=-\cot i ({\bf \Omega}\cdot {\bf n}_{2})$. In this way, we get
\begin{equation}
\dot \varpi = ({\bf \Omega}\cdot {\bf l}) -\cot i ({\bf \Omega}\cdot {\bf n}_{2}),
\label{p3}
\end{equation}  
which is equivalent to the expression (A11) of \cite{IP1}. Now we substitute (\ref{b2}) in (\ref{p3}) and use the facts that $\cos \beta_k
= ({\bf l}\cdot {\bf s}_k)$ and $\cos \delta_k = ({\bf j}\cdot {\bf s}_k)$ to obtain
\begin{equation}
\dot \varpi = \sum_{k=1,2} \left (\Omega_{l}^k+{\Omega_s^{k}\over \sin^2 i}(\cos \beta_k - \cos i \cos \delta_k)\right).
\label{p4}
\end{equation} 
We  use (\ref{e5}) to bring (\ref{p4}) to a more compact form
\begin{equation}
\dot \varpi = \sum_{k=1,2} \left(\Omega_{l}^k+{\Omega_s^{k}\sin \delta_k \cos (\nu_k-\alpha) \over \sin i}\right).
\label{p5}
\end{equation}

Eq. (\ref{p5}) can be brought to an alternative form using the expression (\ref{eq5}) for $\cos \beta_k$
and the expression (\ref{b3}) for $\Omega_{Q,k}$
\begin{align}
&\hspace{-4mm}\dot \varpi=- {\displaystyle{\sum_{k=1,2}}}\Omega_{Q,k} \left(\frac{\left(3\cos^2\beta_k -1 \right)}{2}
+\cot i \cos\beta_k\frac{(\cos\delta_k-  \cos i  \cos\beta_k)}{\sin i}\right).
\label{a15}
\end{align}

{Yet another potentially useful form can be obtained by returning to eq. (\ref{p3})
and making use of the relations
\begin {align}
&  \cot i{\bf \Omega}\cdot{\bf n}_{2} =- {\bf n}_{1}\cdot{\dot {\bf n}}_{2}
= \frac{\cos i}{\sin^2 i}\left(
{\bf j}-\cos i{\bf l}\right)
\cdot{\bf \Omega}.\quad 
\end{align}
Using the fact that
$J{\bf j}=L{\bf l}+\sum_k {\bf S}_k$    
to eliminate ${\bf j}$  in the latter form, we find that
\begin{align}
&  \cot i{\bf \Omega}\cdot{\bf n}_{2} =- {\bf n}_{1}\cdot{\dot {\bf n}}_{2}=  \frac{\cos i}{\sin^2 i} {\displaystyle \sum_{k=1,2}}
\Omega_{s,k}
\left(\left(
\frac{L}{J}-\cos i\right)
\cos\beta_k
+{\displaystyle \sum_{j=1,2}}\frac{S_j{\bf s}_j\cdot{\bf s}_k}{J}
\right).
\label{sub0}
\end{align}
Using eq. (\ref{sub0}) together with eq. (\ref{p3}) then gives
\begin{align}
&  \hspace{-4mm}\dot \varpi=- {\displaystyle{\sum_{k=1,2}}}\Omega_{Q,k} \left(\frac{\left(3\cos^2\beta_k -1 \right)}{2}
 + \frac{\cos i\cos\beta_k}{\sin^2 i}
\left(\left(
\frac{L}{J}-\cos i\right)
\cos\beta_k
+{\displaystyle \sum_{j=1,2}}\frac{S_j{\bf s}_j\cdot{\bf s}_k}{J}
\right)\right).
\label{a16}
\end{align}}

We remark that the first term in the summation in (\ref{a15}), also present in (\ref{a16}),  represents the apsidal advance rate arising from the rotational distortion of the two
 components, this term is denoted as ${\dot \varpi}_\mathrm{R}$
The remaining contribution arises from the non-inertial character of the coordinate frame 
(\ref{en5}) caused
by precession of ${\bf l}$.
We denote this contribution by ${\dot \varpi}_\mathrm{NI}$. Thus both  (\ref{a15}) and (\ref{a16}) 
read
${\dot \varpi}= {\dot \varpi}_\mathrm{R}+{\dot \varpi}_\mathrm{NI}.$

\end{appendix}

\bsp	
\label{lastpage}

\begin{thebibliography}{12}

\bibitem[\protect\citeauthoryear{Albrecht et al.}{2009}]{Albrecht1}Albrecht S., Reffert S., Snellen I. A. G., Winn J. N., 2009, Nature, 461, 373
\bibitem[\protect\citeauthoryear{Albrecht et al.}{2014}]{Albrecht2}Albrecht, S. et al., 2014,  ApJ,  785, Issue 2, 11 
\bibitem[\protect\citeauthoryear{Albrecht, Dawson \& Winn}{2022}]{Albrecht3} Albrecht S. H., Dawson R. I., Winn J. N., 2022, PASP, 134, 082001
\bibitem[\protect\citeauthoryear{Amard et al.}{2019}]{Amard} Amard L., Palacios A., Charbonnel C., Gallet F., Georgy C., Lagarde N., Siess L., 2019, A\&A, 631, A77. 
\bibitem[\protect\citeauthoryear{Barker \& O'Connell}{1975}]{BOC} Barker, B. M.,  O'Connell, R. F.,  1975, Phys. Rev. D, 12, 329
\bibitem[\protect\citeauthoryear{Choi et al.}{2016}]{Choi} Choi J., Dotter A., Conroy C., Cantiello M., Paxton B., Johnson B.~D., 2016, ApJ, 823, 102. 
\bibitem[\protect\citeauthoryear{Claret et al.}{2010}]{claret2010}  Claret A., Torres G.,  Wolf, M., 2010,  A\&A, 515, A4
\bibitem[\protect\citeauthoryear{Dotter}{2016}]{Dotter} Dotter A., 2016, ApJS, 222, 8. 
\bibitem[\protect\citeauthoryear{Ivanov \& Papaloizou}{2021}]{IP} Ivanov P.~B., Papaloizou J.~C.~B., 2021, MNRAS, 500, 3335. (IP) 
\bibitem[\protect\citeauthoryear{Ivanov \& Papaloizou}{2023a}]{IP1} Ivanov P.~B., Papaloizou J.~C.~B., 2023, MNRAS, 526, 3352. (IP1) 
\bibitem[\protect\citeauthoryear{Ivanov \& Papaloizou}{2023b}]{IP2} Ivanov P. B., Papaloizou J.~C.~B., 2023, Astronomy Reports, 67, 912. (IP2) 
\bibitem[\protect\citeauthoryear{Liang, Winn, \& Albrecht}{2022}]{Liang} Liang Y., Winn J.~N., Albrecht S.~H., 2022, ApJ, 927, 114. 
\bibitem[\protect\citeauthoryear{Martynov \& Khaliullin}{1980}]{Mart} Martynov, D. I., Khaliullin, K. F., 1980, ApSS, 71, 177
\bibitem[\protect\citeauthoryear{Ogilvie}{2014}]{Ogilvie2014} Ogilvie G. I., 2014, ARA\&A, 52, 171
\bibitem[\protect\citeauthoryear{Marcussen et al.}{2024}]{Marcussen}  Marcussen, M. L. et al., 2024, arXiv:2408.03072 
\bibitem[\protect\citeauthoryear{Townsend \& Teitler}{2013}]{GYRE} Townsend R.~H.~D., Teitler S.~A., 2013, MNRAS, 435, 3406. 
\bibitem[\protect\citeauthoryear{Paxton et al.}{2011}]{MESA1} Paxton B., Bildsten L., Dotter A., Herwig F., Lesaffre P., Timmes F., 2011, ApJS, 192, 3. 
\bibitem[\protect\citeauthoryear{Paxton et al.}{2013}]{MESA2} Paxton B., Cantiello M., Arras P., Bildsten L., Brown E.~F., Dotter A., Mankovich C., et al., 2013, ApJS, 208, 4. 
\bibitem[\protect\citeauthoryear{Paxton et al.}{2015}]{MESA3} Paxton B., Marchant P., Schwab J., Bauer E.~B., Bildsten L., Cantiello M., Dessart L., et al., 2015, ApJS, 220, 15. 
\bibitem[\protect\citeauthoryear{Paxton et al.}{2018}]{MESA4} Paxton B., Schwab J., Bauer E.~B., Bildsten L., Blinnikov S., Duffell P., Farmer R., et al., 2018, ApJS, 234, 34. 
\bibitem[\protect\citeauthoryear{Paxton et al.}{2019}]{MESA5} Paxton B., Smolec R., Schwab J., Gautschy A., Bildsten L., Cantiello M., Dotter A., et al., 2019, ApJS, 243, 10. 
\bibitem[\protect\citeauthoryear{Philippov \& Rafikov}{2013}]{Philippov} Philippov A.~A., Rafikov R.~R., 2013, ApJ, 768, 112. 
\bibitem[\protect\citeauthoryear{Shakura}{1985}]{Sh} 
Shakura, N. I., 1985, Soviet Astronomy Letters, 11, 224
\bibitem[\protect\citeauthoryear{Sterne}{1939}]{St1939} Sterne T.~E., 1939, MNRAS, 99, 451. 

\bibitem[\protect\citeauthoryear{Ulmer-Moll et al.}{2022}]{UM2022}Ulmer-Moll S. et al., 2022, A\&A, 666, A46
\end{thebibliography}
\end{document}